\documentclass[fleqn,usenatbib]{mnras}

\usepackage{newtxtext,newtxmath}

\usepackage[T1]{fontenc}

\DeclareRobustCommand{\VAN}[3]{#2}
\let\VANthebibliography\thebibliography
\def\thebibliography{\DeclareRobustCommand{\VAN}[3]{##3}\VANthebibliography}

\usepackage{graphicx}	
\usepackage{amsmath}	
\usepackage[dvipsnames]{xcolor} 
\usepackage{ulem}               
\usepackage{upgreek}      

\newcommand{\secRef}[1]{\mbox{Section \ref{sec: #1}}}
\newcommand{\secRefMulti}[1]{\mbox{Sections \ref{sec: #1}}}
\newcommand{\subsecRef}[1]{\mbox{Section \ref{subsec: #1}}}
\newcommand{\subsecRefMulti}[1]{\mbox{Sections \ref{subsec: #1}}}
\newcommand{\appRef}[1]{\mbox{Appendix \ref{app: #1}}}
\newcommand{\appSubsecRef}[1]{\mbox{Appendix \ref{appSubsec: #1}}}
\newcommand{\figRef}[1]{\mbox{Figure \ref{fig: #1}}}
\newcommand{\tabRef}[1]{\mbox{Table \ref{tab: #1}}}
\newcommand{\eqRef}[1]{\mbox{Equation \ref{eq: #1}}}
\newcommand{\eqsRef}[1]{\mbox{Equations \ref{eq: #1}}}
\newcommand{\eqRefMulti}[1]{\mbox{Equations \ref{eq: #1}}}

\mathchardef\shorthyphen="2D                                            
\newcommand{\mSun}{\; {\rm M_\odot}}                            
\newcommand{\mJup}{\; {\rm M_{Jup}}}                            
\newcommand{\mEarth}{\; {\rm M_\oplus}}                         
\newcommand{\mEarthPerMyr}{\; {\rm M_\oplus \; Myr^{-1}}}                    
\newcommand{\rPluto}{\; {\rm R_{Pluto}}}                            
\newcommand{\lSun}{\; {\rm L_\odot}}                            
\newcommand{\au}{\; {\rm au}}                                           
\newcommand{\pc}{\; {\rm pc}}                                           
\newcommand{\m}{\; {\rm m}}                                                     
\newcommand{\cm}{\; {\rm cm}}                                                     
\newcommand{\mm}{\; {\rm mm}}                                                     
\newcommand{\km}{\; {\rm km}}                                           
\newcommand{\um}{\; {\rm \upmu m}}                                        
\newcommand{\yr}{\; {\rm yr}}                                         
\newcommand{\myr}{\; {\rm Myr}}                                         
\newcommand{\mPerS}{\; {\rm m \; s^{-1}}}                       
\newcommand{\kmPerS}{\; {\rm km \; s^{-1}}}                     
\newcommand{\gPerCmCubed}{\; {\rm g \; cm^{-3}}}        
\newcommand{\percent}{\; {\rm per \; cent}}                     
\newcommand{\rad}{\; {\rm radians}}                                         
\newcommand{\HD}{{\rm HD} \;}                                           

\newcommand{\mStar}{m_*}
\newcommand{\lStar}{L_*}
\newcommand{\tStar}{t_*}
\newcommand{\fDisc}{f}
\newcommand{\mDust}{m_{\rm dust}}
\newcommand{\mDisc}{m_{\rm disc}}
\newcommand{\mDiscInitial}{m_{\rm disc, 0}}
\newcommand{\aDisc}{a_{\rm disc}}
\newcommand{\iDisc}{i_{\rm disc}}
\newcommand{\daDisc}{\Delta a_{\rm disc}}
\newcommand{\wDisc}{w_{\rm disc}}
\newcommand{\sMax}{s_{\rm max}}
\newcommand{\sMin}{s_{\rm min}}
\newcommand{\sKm}{s_{\rm km}}

\newcommand{\sDust}{s_{\rm dust}}

\newcommand{\tShep}{t_{\rm shep}}
\newcommand{\mPlt}{m_{\rm P}}

\newcommand{\ds}{{\rm d}s}

\newcommand{\aDeb}{a}
\newcommand{\eDeb}{e}
\newcommand{\iDeb}{i}
\newcommand{\ODeb}{\Omega}
\newcommand{\varpiDeb}{\varpi}
\newcommand{\mP}{m_{\rm p}}

\newcommand{\eP}{e_{\rm p}}

\newcommand{\vCol}{v_{\rm col}}
\newcommand{\eRMS}{\sqrt{\langle e^2 \rangle}}                                         
\newcommand{\eForced}{e_{\rm f}}
\newcommand{\eFree}{e_{\rm p}}
\newcommand{\wForced}{\omega_{\rm f}}
\newcommand{\iFree}{i_{\rm p}}

\newcommand{\yC}{Y_{\rm c}}

\newcommand{\sP}{s_{\rm p}}
\newcommand{\sT}{s_{\rm t}}
\newcommand{\mT}{m_{\rm t}}

\newcommand{\qDStar}{Q_{\rm D}^*}

\newcommand{\qP}{{q_{\rm p}}}
\newcommand{\qG}{{q_{\rm g}}}
\newcommand{\qS}{{q_{\rm s}}}
\newcommand{\bS}{{b_{\rm s}}}
\newcommand{\bG}{{b_{\rm g}}}
\newcommand{\xC}{X_{\rm c}}
\newcommand{\sB}{s_{\rm b}}

\newcommand{\fHahn}{f}
\newcommand{\gHahn}{g}
\newcommand{\mat}[1]{\mathbf{#1}}
\newcommand{\vect}[1]{\mathbf{#1}}

\newcommand{\rhoGas}{\rho_{\rm g}}
\newcommand{\mGas}{m_{\rm g}}
\newcommand{\vRel}{v_{\rm rel}}

\newcommand{\added}[1]{#1}
\newcommand{\removed}[1]{}

\long\def\symbolfootnote[#1]#2{\begingroup%
\def\thefootnote{\fnsymbol{footnote}}\footnote[#1]{#2}\endgroup} 
\def\aj{AJ}
\def\araa{ARA\&A}
\def\apj{ApJ}
\def\apjl{ApJ}
\def\apjs{ApJS}
\def\apss{Ap\&SS}
\def\aap{A\&A}
\def\icarus{Icarus}
\def\jgr{J. Geo. Res.}
\def\mnras{MNRAS}
\def\nat{Nature}

\title[Old Plutos do not dominate Fomalhaut's disc]{Fomalhaut's debris disc is not dominated by primordial Plutos}

\author[T. D. Pearce et al.]{
Tim D. Pearce$^{1}$\thanks{E-mail: tim.pearce@warwick.ac.uk},
Torsten L\"{o}hne$^{2}$,
Alexander V. Krivov$^{2}$
\\
$^1$Department of Physics, University of Warwick, Gibbet Hill Road, Coventry CV4 7AL, UK\\
$^{2}$ Astrophysikalisches Institut und Universit\"{a}tssternwarte, Friedrich-Schiller-Universit\"{a}t Jena, Schillerg\"{a}{\ss}chen 2–3, 07745 Jena,
Germany
}

\date{Accepted XXX. Received YYY; in original form ZZZ}

\pubyear{2023}

\begin{document}
\label{firstpage}
\pagerange{\pageref{firstpage}--\pageref{lastpage}}
\maketitle

\begin{abstract}
A key challenge in debris-disc science is that we do not know the masses of debris discs, nor the sizes of the largest debris bodies. This is because modern observations can only detect objects up to centimetre sizes, whilst larger planetesimals, which dominate disc mass, remain hidden. We must therefore use other arguments, such as dynamics, to indirectly infer disc masses and body sizes. This paper presents a new method, applicable to narrow debris discs like Fomalhaut. We argue that such discs cannot be too massive, nor the largest bodies too large, otherwise they would self-scatter and the disc would be much broader than observed. Using $n$-body dynamics and collisional theory, we demonstrate that the mass of Fomalhaut's disc cannot be dominated by primordial Plutos. Instead, if the mass is dominated by primordial bodies, then they should have radii below ${300^{+80}_{-70}\km}$ (${0.3\pm0.1\rPluto}$) and above ${5^{+20}_{-4}\km}$. Such bodies would each have less than ${1\percent}$ the mass of Pluto. Our conclusions are robust to additional physics, including shepherding planets and collisional damping. Our results provide independent, dynamical support for the idea that the masses of bright debris discs are dominated by objects smaller than Pluto.

\end{abstract}


\section{Introduction}
\label{sec: intro}

Debris discs are circumstellar populations of planetesimals and dust, akin to the Asteroid Belt and Kuiper Belt in our Solar System\footnote{For reviews of debris discs, see \cite{Pearce2024Review, Marino2022, Wyatt2020, Hughes2018, Matthews2014, Krivov2010, Wyatt2008}.}. These discs are detected around ${\sim20\percent}$ of main-sequence stars, through observations of dust in thermal emission or scattered light \citep{Trilling2008, Montesinos2016, Sibthorpe2018}. This dust should only survive for a fraction of the stars' ages before being removed by various physical processes, so it is believed that debris discs also contain larger planetesimals, which continually replenish dust via destructive collisions \citep{Harper1984, Weissman1984}.

A major problem in debris science is that we cannot directly detect these larger planetesimals around other stars. This is because modern instruments are only sensitive to debris smaller than a few centimetres \citep{Hughes2018}, so we do not know the sizes or quantities of the largest bodies in extrasolar debris discs. However, these largest bodies are key to the dynamical and collisional evolution of debris discs, and should dominate their mass. This means that, despite being a fundamental component of planetary systems, \textit{we do not know how massive extrasolar debris discs are}.

Previous attempts to estimate debris-disc masses involved taking the mass of observed dust (typically ${\sim 10^{-2}}$ to ${10^{-1} \mEarth}$; \citealt{Matra2025}), then extrapolating it up to an arbitrary maximum debris-body size (e.g. planetesimals or Plutos) assuming a theoretical size distribution (e.g. \citealt{Dohnanyi1969}). However, these estimates are extremely unreliable, for two reasons. First, we do not know how large the largest debris bodies are; this means we do not know where to stop the extrapolation. Second, there is considerable uncertainty on the form of the size distribution (e.g. \citealt{Najita2022}), and since the extrapolation can occur over a vast size range, even a slight deviation from the assumed size distribution can lead to huge errors in derived masses.

Furthermore, an inescapable issue has emerged known as the `debris-disc mass problem'. If the largest extrasolar debris bodies are hundreds or thousands of kilometres in size, as in our Kuiper Belt (and as predicted by planetesimal-formation models), then many of the brightest extrasolar discs would have extrapolated masses exceeding ${1000 \mEarth}$ \citep{Krivov2018PltmlFormation}. These values are implausibly high, because they exceed the total solid content thought to be inheritable from protoplanetary discs. The most probable solution is that the brightest debris discs comprise bodies no larger than a few kilometres, which is much smaller than the ${1000\km}$ radius of Pluto \citep{Krivov2021}. However, this has not been independently proven.

To proceed, the community needs direct measurements of debris-disc masses and largest-body sizes, which do not rely on extrapolated dust masses. There is no single method for every disc; instead, there are multiple potential methods, each applicable to a subset of discs. Here we present a new method, applicable to narrow discs like Fomalhaut. The idea is that neither the disc mass, nor the size of the largest debris, can be too large, or the disc would scatter itself apart. We use $n$-body simulations and analytics to demonstrate that Fomalhaut's disc cannot be dominated by `primordial' Plutos, i.e. ${\sim1000\km}$ bodies that formed early in the system's lifetime, otherwise the narrow debris disc would have quickly scattered into a broader shape.

In this paper we consider the Fomalhaut system, owing to its well-resolved, narrow debris disc. Fomalhaut A ($\upalpha$ PsA; hereafter Fomalhaut) is a ${440\pm40\myr}$ old A3V star, located at ${7.70 \pm 0.03\pc}$ \citep{vanLeeuwen2007, Mamajek2012}. Its narrow debris disc has a moderate global eccentricity of ${\sim0.1}$, which is well-resolved at multiple wavelengths by the \textit{Hubble Space Telescope}, the Atacama Large Millimeter/submillimeter Array (ALMA), and the \textit{James Webb Space Telescope} \citep{Kalas2013, MacGregor2017, Matra2017, Gaspar2023, Chittidi2025}. The origin of the eccentricity is unclear, because the disc is narrower than expected if it started circular and was later sculpted by an eccentric planet \citep{Faramaz2014, Pearce2014, Kennedy2020}. The system also hosts the enigmatic object \mbox{Fomalhaut b}, on a highly eccentric and potentially disc-crossing trajectory \citep{Kalas2013, Pearce2015Orbits}. The object was first hypothesised as a planet \citep{Kalas2008}, but its lack of thermal emission precluded it from having high mass \citep{Janson2012}, and such an object would disrupt the disc \citep{Beust2014} unless in mean-motion resonance with it \citep{Pearce2021}. However, later imaging showed \mbox{Fomalhaut b} to fade with time, suggesting that it is actually a transient, expanding dust cloud \citep{Gaspar2020}. Regardless the nature of \mbox{Fomalhaut b}, and how disc attained its narrow, eccentric shape, the disc mass and largest-debris size cannot be too large or the disc would scatter itself apart.

The layout of this paper is as follows. \secRef{nBody} describes our $n$-body simulations, showing how discs broaden as a function of disc mass and planetesimal size. \secRef{theoryPrediction} provides an analytic prediction of disc broadening, and \secRef{collisions} combines these dynamical results with collisional theory. \secRef{caveats} validates our findings by considering various \textit{caveats} and additional physics. \secRef{discussionContext} considers our results in the wider context of debris-disc science, and we conclude in \secRef{conclusions}.

\section{$n$-body simulations}
\label{sec: nBody}

We test different combinations of disc mass and largest-debris size for the Fomalhaut disc, and for each combination, assess whether the disc could maintain its narrow shape. We first do this with $n$-body simulations; \subsecRef{nBodySetup} describes the $n$-body setup, \subsecRef{nBodyAnalysis} how the simulations are analysed, and \subsecRef{nBodyResults} the results. We later use dynamical theory and secular-ring simulations to extend these $n$-body results to smaller bodies (\secRef{theoryPrediction} and \appRef{secularSimulations}).

\subsection{Setup of $n$-body simulations}
\label{subsec: nBodySetup}

We perform $n$-body simulations using {\sc rebound} \citep{Rein2012Rebound}. These comprise the Fomalhaut star, and a disc of equal-size, collisional, self-gravitating debris bodies. The debris orbits are initialised to match the observed ALMA disc, using the method in \subsecRef{nBodySetupDisc} and the system parameters in \tabRef{systemPars}.

Each simulation tests a different combination of disc mass $\mDisc$ and debris-body size ${\sMax}$. The smallest disc mass tested is ${0.03 \mEarth}$ (just above the observed $1\mm$ dust mass of ${0.015 \pm 0.010 \mEarth}$; \citealt{MacGregor2017}), and the largest ${10^4 \mEarth}$ (above the ${10^3 \mEarth}$ maximum mass in solids thought inheritable from protoplanetary discs; \citealt{Krivov2021}). Each body's mass is set by equally dividing the disc mass between the number of bodies, where we test between 3 and ${10^4}$ bodies. The body sizes are then calculated from a size-dependent bulk density (\subsecRef{nBodySetupDensity}), and range from ${440\km}$ (${0.37 \rPluto}$) up to unphysically large values above ${10^4 \km}$ for completeness.

\begin{table}
	\centering
	\caption[Caption for LOF]{Parameters of the Fomalhaut system used in this paper. References: [1] \citet{Mamajek2012}; [2] \citet{vanLeeuwen2007}; [3] \citet{MacGregor2017}; [4] Grant Kennedy's \href{http://www.drgmk.com/sdb/}{sdb} library\footnotemark. We use disc parameters from \citet{MacGregor2017}, but more-recent ALMA analyses have since been published (\citealt{Chittidi2025}; \added{\citealt{Lovell2025}}); our values are all within ${3\sigma}$ of the more-recent analyses.}
	\label{tab: systemPars}
	\begin{tabular}{llclc}
		\hline
		Parameter & Name & Value & Unit & Reference\\
        
		\hline
            \multicolumn{4}{c}{\textbf{Star parameters}} \\
            $\mStar$ & Mass & $1.92 \pm 0.02$ & $\mSun$ & [1]\\
            $\lStar$ & Luminosity & $16.63 \pm 0.48$ & $\lSun$ & [1]\\
            $\tStar$ & Age & $440 \pm 40$ & $\myr$ & [1]\\
            $d$ & Distance & $7.70 \pm 0.03$ & $\pc$ & [2]\\

		\hline
            \multicolumn{4}{c}{\textbf{Disc parameters}} \\
            $\aDisc$ & Semimajor-axis median & $136.3 \pm 0.9$ & $\au$ & [3]\\
            $\daDisc$ & Semimajor-axis range & $12.2 \pm 1.6$ & $\au$ & [3]\\
            $\eForced$ & Forced eccentricity & $0.12 \pm 0.01$ & & [3]\\
            $\eFree$ & Proper eccentricity & $0.06 \pm 0.04$ & & [3]\\  
            $\iDisc$ & Midplane inclination to sky & $65.6 \pm 0.3$ & $\deg$ & [3]\\
            PA & Position angle on sky & $337.9 \pm 0.3$ & $\deg$ & [3]\\
            $\wForced$ & Forced argument of pericentre & $22.5 \pm 4.3$ & $\deg$ & [3]\\
            $\mDust$ & Mass in ${1\mm}$ dust grains & $0.015 \pm 0.010$ & $\mEarth$ & [3]\\
            $\fDisc$ & Fractional luminosity & $(5.3 \pm 0.1) \times 10^{-5} $ & & [4]\\
            $h$ & Vertical-aspect ratio & ${0.01 \pm 0.01}$ & & Assumed\\
            $\gamma$ & Surface-density index & -1.5 & & Assumed\\

		\hline
            \multicolumn{4}{c}{\textbf{ALMA-observation parameters}} \\
            $\lambda$ & Observation wavelength & $1.3$ & $\mm$ & [3]\\
            $b_{\rm maj}$ & Synthesized beam major axis & $1.56$ & $\arcsec$ & [3]\\
            $b_{\rm min}$ & Synthesized beam minor axis & $1.15$ & $\arcsec$ & [3]\\
            PA$_{\rm b}$ & Synthesized beam PA & $-87$ & $\deg$ & [3]\\
            
            \hline
	\end{tabular}
\end{table}

\footnotetext{\url{http://www.drgmk.com/sdb/}}
\subsubsection{Initial orbits and positions}
\label{subsec: nBodySetupDisc}

We initialise the debris orbits using the fit to the Fomalhaut disc from ALMA observations (\tabRef{systemPars}; \citealt{MacGregor2017}). This model populates orbits based on free- and forced-eccentricities, which are dynamical concepts describing eccentric discs\footnote{A narrow disc like Fomalhaut can be approximated as having a single forced eccentricity $\eForced$, and a spread of particle eccentricities around $\eForced$ set by the proper eccentricity $\eFree$.}. For each debris body, we randomly draw its semimajor axis $\aDeb$ in the range ${130.2}$ to ${142.4\au}$, assuming the surface density scales as ${\Sigma \propto \aDeb^{-1.5}}$ like the Minimum-Mass Solar Nebula (MMSN; \citealt{Weidenschilling1977, Hayashi1981}). We uniformly draw its proper longitude of pericentre in the range 0 to ${360\deg}$ which, combined with the observed disc's forced eccentricity $\eForced$ and proper eccentricity $\eFree$, yields the debris eccentricity $\eDeb$ and the longitude of pericentre $\varpiDeb$. We use a similar technique to initialise the inclination $\iDeb$ and longitude of ascending node $\ODeb$, assuming a forced inclination of 0 and a proper inclination of ${\iFree = \sqrt{2} h \rad}$, where $h$ is the assumed vertical-aspect ratio of ${0.01\pm0.01}$ (based on \citealt{Boley2012}). Finally, we set the body's initial position on its orbit by uniformly drawing its mean anomaly between 0 and ${360\deg}$. These initial conditions are shown for an example simulation on \figRef{simPlotLowMass}.

\begin{figure*}
	\includegraphics[width=17cm]{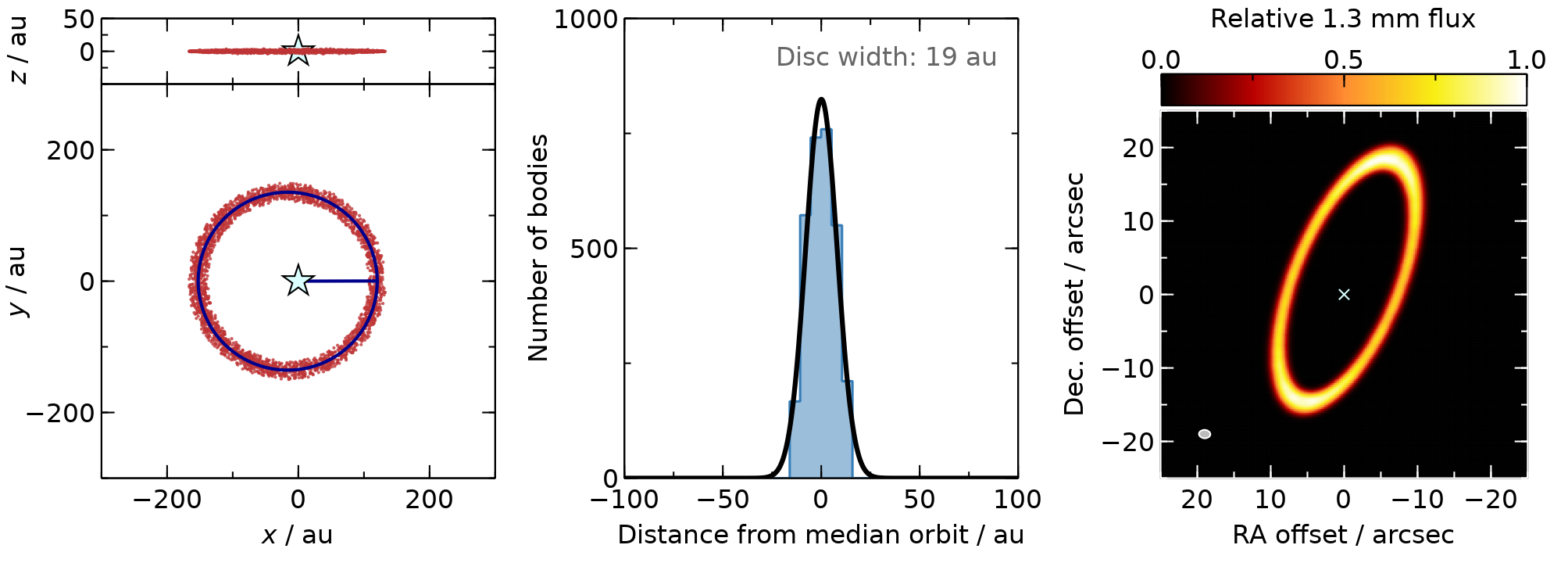}
    \caption{Example setup of an $n$-body simulation. This particular disc has total mass ${1\mEarth}$, and comprises 3000 debris bodies, each of radius ${620\km}$ (${0.52 \rPluto}$). The simulation is shown at time 0, although for these parameters the disc maintains its shape for the whole ${440\myr}$ age of Fomalhaut.
	\textit{Left}: positions of bodies. The star is at the origin, and each brown point is a debris body. The blue ring is the median debris orbit, and the straight blue line shows the pericentre direction of the median orbit.
    \textit{Middle}: distances of debris bodies from the median orbit (blue bars). The black line is a Gaussian fit. 
    \textit{Right}: simulated ${1.3\mm}$ ALMA image of the $n$-body disc. The cross marks the star, and the oval in the bottom left represents the ALMA beam from \citet{MacGregor2017}.}
    \label{fig: simPlotLowMass}
\end{figure*}

\subsubsection{Bulk densities of debris bodies}
\label{subsec: nBodySetupDensity}

Given a debris body's size, we calculate its mass assuming a spherical shape and a size-dependent bulk density. This size dependence arises because larger bodies can gravitationally compact and differentiate. As a proxy, we take the bulk densities of various solid bodies in the Solar System, and fit by eye a simple function for bulk density $\rho$ as a function of body radius $s$ (see \appRef{densities}):

\begin{equation}
  \frac{\rho(s)}{\gPerCmCubed} \approx
    \begin{cases}
      0.8 & \text{if $s < 100\km$;}\\
      0.08\sqrt{\frac{s}{\km}} & \text{else}.
    \end{cases}
    \label{eq: bulkDensityVsSize}
\end{equation}

\noindent We use this to relate debris masses and sizes in our $n$-body simulations.

\subsubsection{Close approaches}
\label{subsec: nBodySetupCollisions}

It is necessary to implement either collisions or softening in self-gravitating $n$-body simulations, to prevent particles making unphysically close approaches and receiving unreasonably large velocity kicks. However, a full collisional-fragmentation prescription is far beyond any $n$-body simulation. Instead we model close approaches using a simple hard-sphere model, where colliding particles rebound off each other to a degree that depends on their coefficient of restitution $\varepsilon$. This coefficient lies between 0 and 1, where 1 is a perfectly elastic collision (bodies rebound in opposite directions with no loss of speed), and 0 is a perfectly inelastic collision (bodies effectively stick together). We use a coefficient of restitution that varies with collision speed $\vCol$ \citep{Bridges1984}, as often used in simulations of planetary rings (e.g. \citealt{Ohtsuki1999, Salo2018}):

\begin{equation}
	\varepsilon = \min\left[0.109 \left(\frac{\vCol}{\mPerS} \right)^{-0.234}, 1 \right].
    \label{eq: coefficientOfRestitution}
\end{equation}

For typical collision speeds in debris discs (${\gtrsim 10\mPerS}$; \citealt{Costa2024}), \eqRef{coefficientOfRestitution} yields a coefficient of restitution close to 0 (inelastic collisions). Our $n$-body simulations therefore incorporate maximal collisional damping between equal-size particles, and if such collisions occurred, our simulated particles would effectively stick together. However, we find damping to be negligible in our simulations, owing to very low collision frequencies.

In reality, collisional debris follows a size distribution, with smaller bodies vastly outnumbering larger ones \citep{Dohnanyi1969}. In certain regimes, collisions between unequal-sized bodies could significantly damp the larger object \citep{Jankovic2024}, which is unavoidably neglected in our single-size $n$-body simulations. However, in \subsecRef{caveatsCollisionsDamping} we will show that realistic collisional damping is negligible in our parameter space of interest, which justifies its omission from our simulations.

\subsubsection{Running $n$-body simulations}
\label{subsec: nBodySetupIntegration}

Each simulation uses one of two $n$-body prescriptions, depending on the number of bodies $n$. For smaller numbers of particles (${n < 3000}$), simple $n$-body gravity is used, where each body feels the exact gravitational force from every other body; in this case the computational requirements scale as $n^2$. These simulations use {\sc rebound}'s {\sc mercurius} integrator \citep{Rein2019Mercurius}, which employs the fixed-timestep {\sc whfast} integrator when bodies are far apart, and switches to the {\sc ias15} adaptive-stepsize integrator for close approaches. For these simulations we set {\sc whfast}'s timestep to be ${10^{-3} P_{\rm 0, min}}$, where $P_{\rm 0, min}$ is the orbital period of the innermost particle at the start of the simulation. We also constrain the {\sc ias15} adaptive timestep to not drop below ${10^{-9} P_{\rm 0, min}}$. Both timesteps are chosen to ensure high accuracy in scattering interactions, exceeding the accuracy of previous works (e.g. \citealt{Pearce2024Edges}). Collisions are detected using the {\sc line} algorithm.

Conversely, for larger numbers of particles (${n \geq 3000}$), we use tree gravity. The physical space is divided into cells, and forces between distant particles are approximated as the total of all particles in a cell. If particles are too close, then cells are subdivided until sufficient accuracy is achieved \citep{Rein2012Rebound}. The computational requirements scale as ${n \ln(n)}$, so tree gravity is considerably faster than $n$-body gravity for large numbers of particles. For tree simulations we use {\sc rebound}'s {\sc leapfrog} fixed-timestep integrator, with a timestep of ${10^{-4} P_{\rm 0, min}}$. The box size is ${3000\au}$, and particles leaving this box are removed from the simulation. The parameter \texttt{opening\_angle2}, which sets the tree-code accuracy, is set to ${1.5}$ as in the {\sc rebound} tree-code example\footnote{The opening angle is the ratio of a cell's width to its distance from a particle. The parameter \texttt{opening\_angle2} is the square of the critical opening angle, which sets how cells are divided when calculating the force on a particle. If a cell's opening angle is larger than the critical value, then the cell is subdivided until the opening angles are sufficiently small.}\textsuperscript{, }\footnote{\url{rebound.readthedocs.io/en/latest/c_examples/selfgravity_disc/}}. These computational parameters are chosen as a balance of speed and accuracy based on convergence testing, and verified against $n$-body gravity in \appRef{treeCodeTests}. Collisions are detected using the {\sc line-tree} algorithm.

All simulations are either run for the ${440\myr}$ age of Fomalhaut, or until the disc width increases by a factor of 5, whichever occurs first (disc widths are calculated using the method in \subsecRef{nBodyAnalysis}). This time cutoff is justified because we find that a disc, once broadened, will not coalesce back into a narrow ring through self-gravity; we show this in \appRef{secularSimulations}. Furthermore, in \subsecRef{caveatsCollisionsDamping} we will show that realistic collisional damping is unlikely to change this.

\subsection{Analysis of $n$-body simulations}
\label{subsec: nBodyAnalysis}

To quantify how the Fomalhaut disc broadens, we fit the belt width at the end of the simulation, and compare it to the initial width. We describe the fitting process here.

Since the disc is eccentric, we cannot derive its width using an axisymmetric radial profile. Instead, we first calculate the `median' debris orbit. We define this orbit's semimajor axis as the median \added{semimajor axis} of the debris bodies, and set its inclination to zero relative to the disc midplane. Its eccentricity and longitude of pericentre are set by finding the median eccentricity vector of the debris bodies: ${(e\cos\varpi, \; e\sin\varpi)}$. An example of this median orbit is the blue ellipse on the left panel \figRef{simPlotLowMass}.

We then calculate the distance between each debris body and the median orbit, by sampling the median orbit and finding the point closest to the debris body\footnote{Minimising the distance between a point and an ellipse is a problem with no general analytic solution. Our brute-force method proved sufficiently fast and reliable.}. The middle panel of \figRef{simPlotLowMass} shows a histogram of these distance for that simulation. Finally, we fit the histogram with a Gaussian (line on the middle panel of \figRef{simPlotLowMass}), and define the disc width as the Gaussian's full-width-half-maximum.

To reduce the effect of random noise, if there are fewer than 100 particles in a simulation, then before calculating the disc width we populate each orbit with 100 pseudo-particles at uniformly randomised mean anomalies. We also run each of those simulations at least 10 times with different seeds, then take the median width.

The right panel of \figRef{simPlotLowMass} shows a simulated ${1.3\mm}$ ALMA observation of the example $n$-body disc, for illustration only. To generate this image, we took each debris body and treated it like a collection of ${1.3\mm}$ dust grains. We populated each orbit with 100 pseudo-grains at uniformly random mean anomalies, then calculated each pseudo-grain's temperature as proportional to $r^{-1/2}$ (where $r$ is distance from the star). We then used this temperature to calculate the grain's relative blackbody emission. Finally we rotated the simulation to the system orientation on the sky, and convolved the emission at each point with a 2-dimensional rotated Gaussian representing the ALMA beam of \citet{MacGregor2017}. These images are not used in our quantitative analyses.

\subsection{Results of $n$-body simulations}
\label{subsec: nBodyResults}

Some discs in our $n$-body simulations maintain their narrow, eccentric shape for the ${440\myr}$ age of Fomalhaut. This includes the example on \figRef{simPlotLowMass}. Other discs scatter themselves into much broader structures that are incompatible with observations, like that on \figRef{simPlotHighMass}. The broadening time varies between simulations, but in many simulations it is orders-of-magnitude shorter than the system age.

\begin{figure*} 
    \includegraphics[width=17cm]{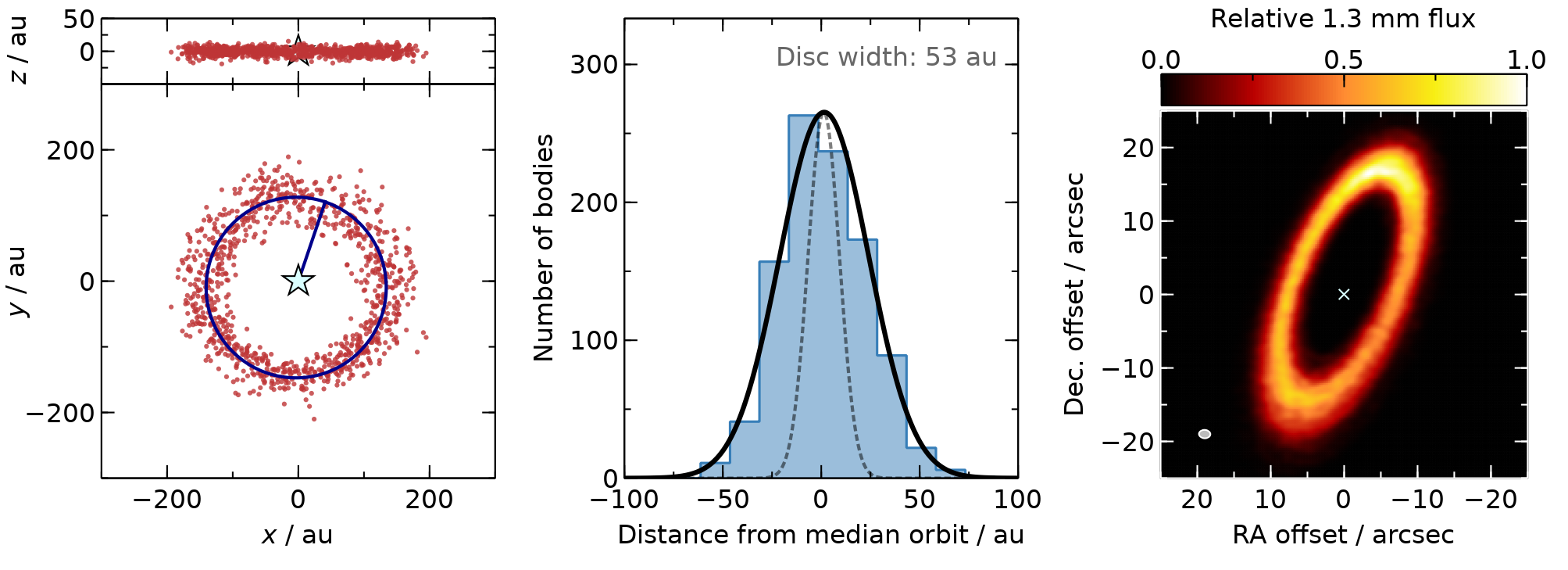}
    \caption{Example $n$-body simulation where the disc significantly broadens via self-scattering. This disc has a total mass of ${30\mEarth}$, and comprises 1000 debris bodies each of radius ${2200\km}$ (${1.9 \rPluto}$). This simulation started in a similar configuration to \figRef{simPlotLowMass}, and is shown after ${160\myr}$; by this time the disc has almost tripled in width, and is incompatible with observations. Definitions and most axis scales are the same as \figRef{simPlotLowMass}, although the histogram vertical axis has been rescaled due to the different number of bodies, and the flux on the right panel has a different normalisation. The dotted line on the middle panel shows the initial-disc profile.}
    \label{fig: simPlotHighMass}
\end{figure*}

\figRef{discMassSMaxAllowedRegions} shows the results of all $n$-body simulations. Each cross marks a simulation, and the colourmap shows how much the simulated discs broaden. There is a clear divide between behaviours; if a disc is too massive, or the bodies too large, then the disc scatters into a broader structure (upper-right region of \figRef{discMassSMaxAllowedRegions}). Conversely, if the disc mass and/or body size is small, the disc maintains its narrow shape across the system age. Disc mass alone does not decide this behaviour; for a given disc mass, discs with smaller bodies are better able to remain narrow. In \secRef{theoryPrediction} we will verify these results using dynamical theory.

\begin{figure}
	\includegraphics[width=8cm]{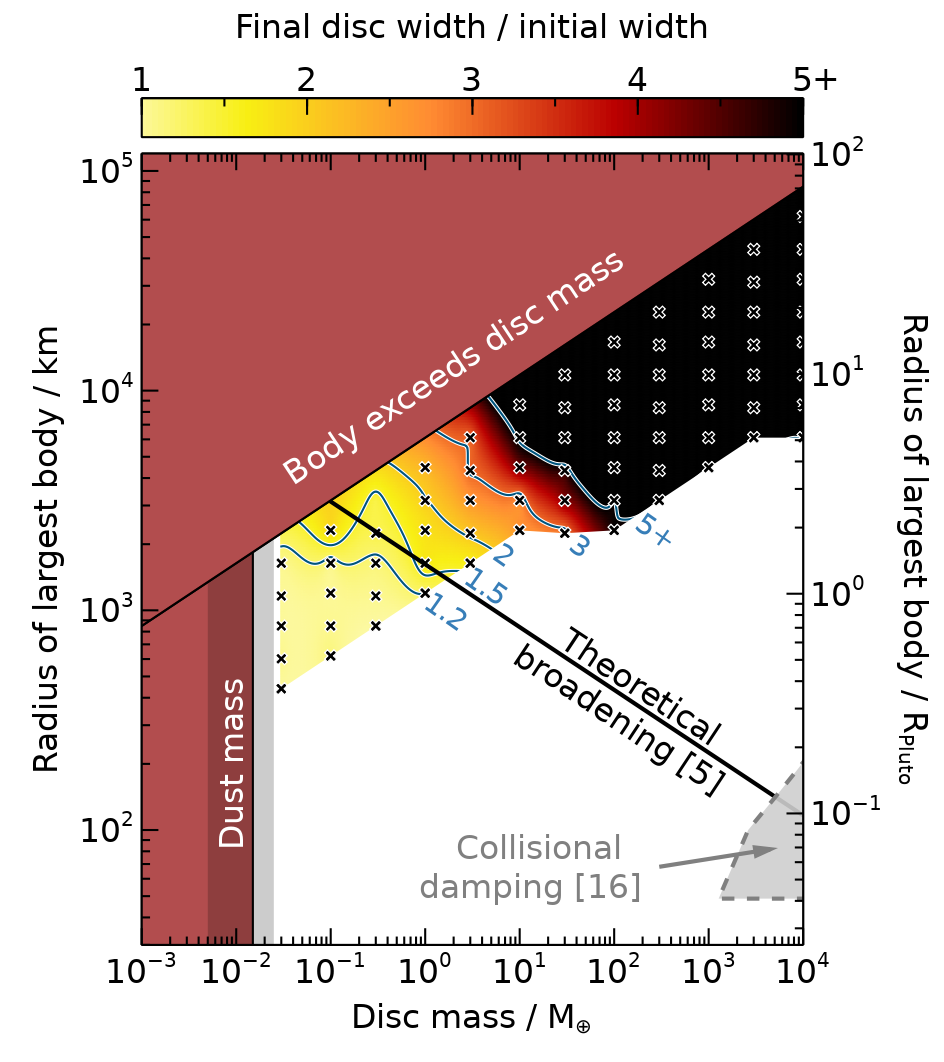}
    \caption{Expected broadening of Fomalhaut's debris disc, as a function of disc mass and the size of the largest debris bodies. Crosses mark $n$-body simulations. The colour scale and contour lines show the interpolated final width of those simulated discs, divided by their initial widths. Discs that are too massive, or made up of bodies that are too large, readily scatter into broader structures (upper-right region). Red blocked areas are unphysical; the total disc mass must be larger than that in observed millimetre dust (left), and one body cannot be more massive than the whole disc (top). The `Theoretical broadening' line is our prediction, above which a narrow planetesimal disc would significantly widen through self scattering (\eqRef{maxMDiscBeforeScatter}). The `Damping' area is where realistic collisional damping could cause the disc to narrow (\eqRef{minMDiscToDamp}). Numbers in square brackets denote equations, and the shading around the `Dust mass' line is the observational uncertainty (\tabRef{systemPars}).
    }
    \label{fig: discMassSMaxAllowedRegions}
\end{figure}

Our initial-disc widths are ${19\au}$, which are wider than the synthesised ALMA beam; the beam has major axis ${1.56\arcsec \times 7.7 \pc = 12\au}$, and minor axis  ${1.15\arcsec \times 7.7 \pc = 8.9\au}$ \citep{MacGregor2017}. This means that the initial belt, which matches observations, is resolved by 1.6 to 2.1 beams at its ansae. We conservatively argue that we can rule out simulations where the disc broadens by more than a factor of 1.5, because this difference would be detectable in ALMA observations. This suggests that we can already rule out largest-debris sizes above ${\sim4\rPluto}$, based on \figRef{discMassSMaxAllowedRegions}, because larger bodies would significantly broaden the disc. However, we will produce tighter constraints in \secRef{collisions}, by combining our dynamical arguments with collisional theory.

These results assume that the initial disc had the same width as the observed disc. This is valid for two reasons. First, the disc is unlikely to have be born wider and then narrowed, for the reasons given in \appRef{secularSimulations}. Second, if the initial disc were narrower, then the initial number density of particles would have been higher, so scattering would be more efficient and the disc would broaden even more. In \secRef{theoryPrediction} we will show that the eccentricity of self-scattering bodies grows faster for narrower discs, so narrow initial discs would broaden faster. This means that our upper limits on disc mass and largest-body size still apply, even if the initial disc were narrower.

\section{Theoretical prediction of disc broadening}
\label{sec: theoryPrediction}

We now use dynamical theory to predict the disc masses and debris sizes required to significantly broaden a narrow disc. This verifies our $n$-body simulations, and provides general predictions applicable to other systems.

\cite{Ida1993} quantify the viscous-stirring timescale for a disc of self-stirring planetesimals. They start with an axisymmetric disc of equal-mass planetesimals, with low initial eccentricities and inclinations, and quantify how the root-mean-square eccentricity $\eRMS$ grows with time $t$. Combining their \mbox{Equations 4.1} and 4.2, and deriving their surface-number density of planetesimals using the disc width $\wDisc$, yields

\begin{equation}
  \eRMS = \left(\frac{2}{3}\frac{C_{\rm e} \sqrt{G} s^3\rho \mDisc}{\mStar^{3/2} \sqrt{\aDisc} \wDisc} t \right)^\frac{1}{4}
    \label{eq: eRMS}
\end{equation}

\noindent or equivalently

\begin{multline}
  \eRMS = 1.261\times10^{-4} \left(\frac{C_{\rm e}}{40}\right)^\frac{1}{4} \left(\frac{s}{\km}\right)^\frac{3}{4} \left(\frac{\rho}{\gPerCmCubed}\right)^\frac{1}{4} \left(\frac{\mDisc}{\mEarth}\right)^\frac{1}{4}\\
  \times  \left(\frac{\aDisc}{\au}\right)^{-\frac{1}{8}} \left(\frac{\wDisc}{\au}\right)^{-\frac{1}{4}} \left(\frac{\mStar}{\mSun}\right)^{-\frac{3}{8}} \left(\frac{t}{\myr}\right)^\frac{1}{4},
    \label{eq: eRMSUnits}
\end{multline}

\noindent where ${C_{\rm e} \sim 40}$ is a numerical constant \citep{Ida1993}\footnote{\eqRefMulti{eRMS} and \ref{eq: eRMSUnits} are comparable to \mbox{Equation 11} in \cite{Matra2019}, except that ours apply to planetesimal-planetesimal scattering, rather than planet-planetesimal scattering (so are smaller than \citealt{Matra2019} by a factor of $2^{1/4}$), and ours apply to eccentricity rather than inclination (for inclination, substitute $e$ for $i$ and $C_{\rm e}$ for ${C_{\rm i} \sim 2}$ in \eqRefMulti{eRMS} and \ref{eq: eRMSUnits}; \citealt{Ida1993}).}.

Applying this to Fomalhaut: a ring of particles with semimajor axis $\aDisc$, forced eccentricity $\eForced$ and proper eccentricity $\eFree$ has a width ${\wDisc = 2\aDisc\eFree}$. Hence Fomalhaut's disc would be significantly wider if the root-mean-square eccentricities from self scattering were greater than ${\wDisc/(2\aDisc)}$. So rearranging \eqRef{eRMSUnits} yields the maximum initial mass the narrow disc could have, as a function of largest-debris size, without significantly broadening within the age of the star:

\begin{multline}
  \left(\frac{\mDiscInitial}{\mEarth}\right) \lesssim 2.5 \times 10^{14} \; 
  \left(\frac{C_{\rm e}}{40}\right)^{-1} \left(\frac{s}{\km}\right)^{-3} \left[\frac{\rho(s)}{\gPerCmCubed}\right]^{-1}
  \\
  \times \left(\frac{\aDisc}{\au}\right)^{-\frac{7}{2}} \left(\frac{\daDisc}{\au}\right)^5 \left(\frac{\mStar}{\mSun}\right)^{\frac{3}{2}} \left(\frac{\tStar}{\myr}\right)^{-1}.
    \label{eq: maxMDiscBeforeScatter}
\end{multline}

\noindent Here we used the semimajor-axis range $\daDisc$ as a proxy for the initial-disc width $\wDisc$. In reality the semimajor axes would also spread, which would broaden the ring even more; hence \eqRef{maxMDiscBeforeScatter} is a firm upper limit on the maximum initial mass a narrow disc could have without significantly broadening.

We plot \eqRef{maxMDiscBeforeScatter} as the solid black line on \figRef{discMassSMaxAllowedRegions}, using the size-dependent bulk density from \eqRef{bulkDensityVsSize}. The line shows good agreement with our simulations; simulations below the line maintain their narrow shapes, whilst those above the line undergo broaden by a factor of at least 1.3. \eqRef{maxMDiscBeforeScatter} is therefore a good predictor for the maximum mass a narrow planetesimal disc can have without scattering itself apart.

\section{Combining dynamics with collisions}
\label{sec: collisions}

\secRefMulti{nBody} and \ref{sec: theoryPrediction} gave purely dynamical constraints on disc mass and largest-body size. However, we get more-powerful constraints by combining dynamics with collisional theory. Here we use several collisional arguments to further constrain the allowed parameter space, with the resulting bounds shown on  \figRef{discMassSMaxAllowedRegionsCollisions}. Using this, we show that the mass of Fomalhaut's debris disc cannot be dominated by primordial Plutos. The arguments are as follows.

\begin{figure}
    \includegraphics[width=8cm]{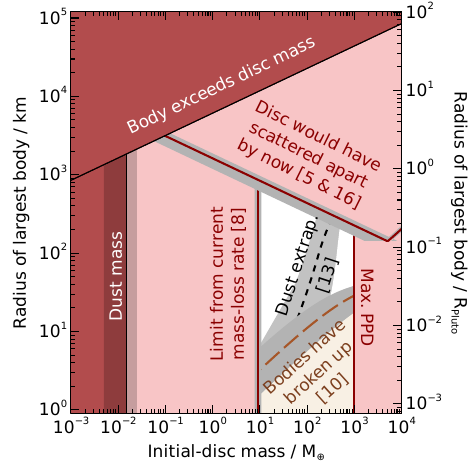}
    \caption{Combined dynamical and collisional constraints on the Fomalhaut debris disc. Red shaded regions are ruled out by physical arguments. The numbers in square brackets denote the corresponding equation numbers. The beige region below \mbox{Line \ref{eq: tau0FromM0AndSMax}} is not necessarily excluded, but shows where most of the largest primordial bodies would have collisionally broken up by now. \mbox{Line \ref{eq: massIntegral}} is an extrapolation from observed dust; if debris follows the 3-powerlaw size distribution of \citet{Lohne2008}, then the total disc mass and largest-body size lie along this line. `Max. PPD' denotes the ${10^3 \mEarth}$ maximum mass in solids thought inheritable from protoplanetary discs \citep{Krivov2021}. The horizontal axis is the disc's initial mass, but for the top $\sim$two-thirds of the plot (radii ${>48\km}$), the initial-disc mass equals the present-day mass, because the largest bodies have not yet started colliding. Shading around lines denotes ${1\sigma}$ uncertainties, propagated from observational uncertainties in \tabRef{systemPars}. Note that the vertical-axis scale is larger than \figRef{discMassSMaxAllowedRegions}.
    }
    \label{fig: discMassSMaxAllowedRegionsCollisions}
\end{figure}

\subsection{Lower bound on disc mass from mass-loss rate}
\label{subsec: collisionsMinDiscMass}

The disc mass must be larger than the observed-dust mass of ${0.015 \pm 0.010 \mEarth}$ \citep{MacGregor2017}, but a more-stringent constraint comes from collisional evolution. Following \cite{Krivov2021}, a debris disc in collisional cascade, where bodies are predominantly destroyed by collisions with similar-sized bodies, has mass decreasing with time $t$ as

\begin{equation}
  \mDisc (t) = \frac{\mDiscInitial}{1 + t / \tau_0} .
    \label{eq: discMassVsTime}
\end{equation}

\noindent Here $\mDiscInitial$ is the initial-disc mass, and $\tau_0$ is the collisional lifetime of the largest bodies at the initial epoch \citep{Wyatt2007AStars, Lohne2008}. Differentiating this, and setting $t$ to the star age $\tStar$, yields

\begin{equation}
 \mDiscInitial = |\dot{m}_{\rm disc} (\tStar)| \tau_0 \left(1 + \tStar / \tau_0 \right)^2.
   \label{eq: discMassLossRate}
\end{equation}

\noindent Here ${|\dot{m}_{\rm disc} (\tStar)|}$ is the current mass-loss rate, which can be estimated from observations (see below). The value $\tau_0$ depends on the disc parameters, disc mass, and the largest-body size, but by differentiating \eqRef{discMassLossRate} with respect to ${\tau_0}$, we see that the equation is minimised if ${\tau_0 = \tStar}$. Hence a lower bound on the initial-disc mass is

\begin{equation}
  \mDiscInitial \gtrsim 4 |\dot{m}_{\rm disc}| \tStar.
    \label{eq: minInitialDiscMassFromMassLossRate}
\end{equation}

We now apply this to Fomalhaut, using the parameters and uncertainties from \tabRef{systemPars}. The mass-loss rate is estimated using \mbox{Equation 21} in \citet{Matra2017}, yielding ${|\dot{m}_{\rm disc}| = (5.3 \pm 0.7) \times 10^{-3} \mEarthPerMyr}$. Inserting this into \eqRef{minInitialDiscMassFromMassLossRate} yields a lower bound on the initial-disc mass of ${9 \pm 2 \mEarth}$. This limit is shown on \figRef{discMassSMaxAllowedRegionsCollisions}, and lies comfortably between equivalent lower limits of ${3.6 \mEarth}$ from \cite{Krivov2021} and ${19 \mEarth}$ from \cite{Matra2017}\footnote{${3.6 \mEarth}$ comes from substituting the \citet{Krivov2021} lower bound on the \textit{current} mass (${1.8 \mEarth}$) into \eqRef{discMassVsTime} with ${t=\tau_0}$. ${19 \mEarth}$ comes from substituting the \citet{Matra2017} estimate of ${|\dot{m}_{\rm disc}| = 1.1 \times 10^{-2} \mEarthPerMyr}$ into \eqRef{minInitialDiscMassFromMassLossRate}.}.

By comparing \eqsRef{maxMDiscBeforeScatter} and \ref{eq: minInitialDiscMassFromMassLossRate}, we see that the largest primordial bodies cannot be Plutos, because in that case the minimum-mass disc would have scattered apart by now. From the intercept of these lines on \figRef{discMassSMaxAllowedRegionsCollisions}, the largest primordial body must have a radius below ${0.7^{+0.1}_{-0.2} \rPluto}$ (${800\pm100\km}$).

\eqRef{minInitialDiscMassFromMassLossRate} is a lower bound on the initial-disc mass, but if $\tau_0$ is large enough, then the largest bodies will not yet have started colliding. In this case the present-day disc mass will be the same as its initial mass, and the lower bound of ${9 \pm 2 \mEarth}$ from \eqRef{minInitialDiscMassFromMassLossRate} also applies to the present-day disc. As we will show in \subsecRef{collisionsLifetimeOfLargestBodies}, this is true regardless of disc mass if the largest body has radius ${\gtrsim48\km}$ (if made of basalt) or ${\gtrsim250\km}$ (if made of ice). Also, whilst the lower bound from \eqRef{minInitialDiscMassFromMassLossRate} includes an implicit assumption about the dust-size distribution, in \subsecRef{caveatsSizeDist} we show the results are robust for reasonable assumptions.

\subsection{Collisional lifetimes of the largest bodies}
\label{subsec: collisionsLifetimeOfLargestBodies}

We can also constrain whether the largest primordial bodies would survive until today. A large body's collisional lifetime increases with body size (because larger bodies are stronger), and decreases with disc mass (because fewer objects means fewer collisions). So for a given initial-disc mass, there is a lower limit on the size of a primordial body for it to survive until the present day. 

We use the collisional prescription of \cite{Lohne2008}. This assumes that debris follows a triple-powerlaw size distribution for radii between $\sMin$ and $\sMax$, where $n(s)\ds$ is the number of bodies with radii between $s$ and $s + \ds$, and

\begin{equation}
    n(s) \propto
    \begin{cases}
      s^{2 - 3\qS} & \text{if}\ \sMin \leq s < \sB, \\
      s^{2 - 3\qG} & \text{if}\ \sB \leq s < \sKm, \\
      s^{2 - 3\qP} & \text{if}\ \sKm \leq s < \sMax, \\
      0 & \text{else}.
    \end{cases}
    \label{eq: sizeDist}
\end{equation}

\noindent Here $\sKm$ is the radius of the largest \textit{colliding} body, $\sMax$ the radius of the largest \textit{primordial} body (which, if ${\sKm < \sMax}$, has not yet entered the collisional cascade), and $\sB$ the radius of the weakest body (around the size where bodies transition from being held together by material strength to gravity). Following \cite{Lohne2008}, we use ${\qS = 1.89}$, ${\qG = 1.68}$ and ${\qP = 1.87}$; these correspond to ${n(s) \propto s^{-3.67}}$, $s^{-3.04}$ and $s^{-3.61}$ in the three respective regimes. We discuss the impact of uncertainties on these values in \subsecRef{caveatsSizeDist}.

\mbox{Equation 42} in \cite{Lohne2008} gives $\tau_0$, the collisional lifetime of the largest body at the initial epoch, as a function of body size $\sMax$, initial-disc mass $\mDiscInitial$, and other parameters\footnote{\citealt{Pearce2024Edges} note that \mbox{Equation 42} in \cite{Lohne2008} has an erroneous index of -1 around the square bracket, which we omit here. This is unimportant provided ${\qP<2}$ and ${\sMin\ll\sMax}$.}. However, $\tau_0$ is not necessarily the \textit{actual} collisional lifetime, because over time the disc mass decreases, and hence a body's collisional lifetime increases. So to get a lower limit on the largest primordial body that could survive until today, we set ${\tau_0 = \tStar}$ and the disc mass to ${\mDiscInitial/2}$; this is the present-day disc mass if ${\tau_0 = \tStar}$ (\eqRef{discMassVsTime}). We do this because a primordial body that were too small to survive until today with the collisionally reduced disc mass would be even less likely to survive in reality, because the initial-disc mass was even larger. Hence the maximum initial-disc mass that allows the majority of the largest bodies to survive until the present day is

\begin{multline}
    \mDiscInitial(\sMax) < 2\times\frac{16 \pi \rho(\sMax)}{3 \tStar} \frac{\sMax \aDisc^{5/2} \daDisc}{\sqrt{\mathcal{G} m_*}} \\
    \times \frac{\qG -5/3}{2-\qP} \left[ 1 - \left(\frac{\sMin}{\sMax} \right)^{6-3\qP}\right] \frac{i}{f(e,i) G(\qG, \sMax, \aDisc)}.
    \label{eq: tau0FromM0AndSMax}
\end{multline}

\noindent Here $\mathcal{G}$ is the gravitational constant, $e$ and $i$ are the typical relative eccentricities and inclinations of bodies respectively, and the functions ${f(e,i)}$ and ${G(q, s, r)}$ are defined in \appRef{collisionFunctions}.

\eqRef{tau0FromM0AndSMax} depends on the material strength ${\qDStar}$, which is contained in ${G(q, s, r)}$. To make a conservative estimate of the maximum disc mass, we assume debris is composed of strong basalt; weaker materials would break apart sooner, so the disc would have to be even less massive for those bodies to survive. Following \cite{Lohne2012}, \cite{Schuppler2016} and \cite{Krivov2018PltmlFormation}, we use a size-dependent fragmentation energy of

\begin{equation}
    \qDStar(s,r) = \left(\frac{v_{\rm col}}{v_0} \right)^{1/2}
    \left[A_{\rm s} \left(\frac{s}{1\m}\right)^{-3\bS}
    + A_{\rm g} \left(\frac{s}{1\km}\right)^{3\bG} \right],
    \label{eq: qDStar}
\end{equation}

\noindent where ${v_0=3\kmPerS}$, ${A_{\rm s}=A_{\rm g}=5\times10^6 \; {\rm erg \; g^{-1}}}$, ${b_{\rm s}=0.12}$ and ${b_{\rm g}=0.46}$. For this material, $\qDStar$ is minimised for bodies with radii equal to the breaking radius ${\sB = 110\m}$. The collision speed $v_{\rm col}$ is

\begin{equation}
    v_{\rm col} = \sqrt{\frac{\mathcal{G} m_*}{\aDisc}} f(e, I).
    \label{eq: vCol}
\end{equation}


We now apply this to Fomalhaut: we set ${e=\eP}$ and ${i=\sqrt{2}h}$, which yields typical collision speeds of ${\vCol \sim 240 \pm 10 \mPerS}$ in the disc. We arbitrarily set ${\sMin = 1\um}$, and use the size-dependent bulk density ${\rho(s)}$ from \eqRef{bulkDensityVsSize}. Using these, \eqRef{tau0FromM0AndSMax} yields the maximum initial-disc mass as a function of body size, as shown on \figRef{discMassSMaxAllowedRegionsCollisions}.

This line asymptotically tends towards ${48\km}$ for high disc masses. This is because, for basalt bodies, anything with radius larger than ${\sim48\km}$ is strong enough to never break up in the Fomalhaut disc (i.e. ${\tau_0 \rightarrow \infty}$)\footnote{The minimum debris size to never break apart is given by \mbox{Equation 31} in \cite{Pearce2024Edges}.}. From \eqRef{discMassVsTime}, this means that for the upper $\sim$two-thirds of \figRef{discMassSMaxAllowedRegionsCollisions} (radii ${\gtrsim48\km}$), the current disc mass equals its initial mass. We also note that this critical size is well below the sizes tested in our $n$-body simulations; hence those bodies would still exist today, and we are safe to omit collisional erosion from our $n$-body simulations\footnote{If the disc lost mass through collisional erosion, then dynamical broadening would be reduced. However, this would require the disc parameters to lie in the bottom-right of \figRef{discMassSMaxAllowedRegionsCollisions} (below \eqRef{tau0FromM0AndSMax}), in which case dynamical broadening is negligible anyway. Hence mass loss does not affect our dynamical conclusions.}.

Whilst bodies larger than ${\sim48\km}$ would survive indefinitely, smaller primordial bodies could have broken up by now, if the disc were massive enough. This scenario is not necessarily precluded, but if the largest primordial bodies have since broken up, then the primordial disc must have been more massive and brighter than today (\eqRef{discMassVsTime}; \citealt{Wyatt2007AStars, Lohne2008}). Combining \eqRef{tau0FromM0AndSMax} with the minimum disc mass from \eqRef{minInitialDiscMassFromMassLossRate}\removed{,} we see that\added{,} for a primordial body to survive until today, it would require a radius larger than ${3^{+2}_{-1} \km}$. Bodies composed of weaker materials would have stricter limits; icy bodies could have ${A_{\rm s}}$ and ${A_{\rm g}}$ reduced by a factor of 10 in \eqRef{qDStar} \citep{Krivov2005}, in which case icy primordial bodies would require radii larger than at least ${7^{+4}_{-2}\km}$ to survive until the present day, and larger for higher disc masses. Similarly, any primordial icy body with radius above ${\sim250\km}$ could survive indefinitely, for any mass of the Fomalhaut disc.

\subsection{Extrapolating dust mass}
\label{subsec: collisionsExtrapolationFromDust}


The final constraints come from dust mass. If we know the form of the size distribution, then we can estimate the total mass in bodies up to a given size, by extrapolating the mass in observed dust. This technique cannot uniquely constrain disc mass, because we do not know the size of the largest body and hence how far to extrapolate, but it can relate disc mass and largest-body size.

Here we constrain the largest-body size as a function of initial-disc mass $\mDiscInitial$. We do this numerically; for a given initial-disc mass, we interpolate to find the corresponding largest-body size $\sMax$. First, we normalise  the present-day size distribution (\eqRef{sizeDist}) at the smallest sizes. The total mass of spherical bodies with radii between $s_1$ and $s_2$ is

\begin{equation}
    m_{s_1 \rightarrow s_2} = \frac{4\pi}{3}\int^{s_2}_{s_1} n(s) \rho(s) s^3 \text{d}s,
    \label{eq: massIntegral}
\end{equation}

\noindent so we normalise the size distribution from $\sMin$ to $\sB$ by setting ${m_{\sMin \rightarrow \sDust} = \mDust}$. 

We then normalise the second part of the size distribution, from $\sB$ to $\sKm$, by assuming the distribution is continuous at $\sB$. The upper end of this part is $\sKm$, the radius of the current largest colliding body. This can be found from \mbox{Equation 31} in \cite{Lohne2008}, which gives the collisional lifetime of an object of radius $s$:

\begin{multline}
    \tau = \frac{16\pi\rho(s)}{3\mDiscInitial} \left(\frac{s}{\sMax}\right)^{3\qP-5} \frac{\sMax\aDisc^{5/2}\daDisc}{\sqrt{\mathcal{G}\mStar}} \\ \times \frac{\qP - 5/3}{2-\qP} \left[ 1-\left(\frac{\sMin}{\sMax}\right)^{6-3\qP}\right] \frac{i}{f(e,i) G(\qP, s, \aDisc)}.
    \label{eq: tauOfGeneralS}
\end{multline}

\noindent Solving this for $s$ at ${\tau = \tStar}$ yields $\sKm$. Bodies larger than $\sKm$ have not yet started colliding, so the size distribution has the primordial slope from $\sKm$ to $\sMax$. Again, we normalise this part by assuming the distribution is continuous at $\sKm$.

Having normalised the \textit{present-day} size distribution at $\sKm$, we can finally calculate $\sMax$ as a function of $\mDiscInitial$ by integrating along the \textit{primordial} size distribution (i.e. ${\mDiscInitial = m_{\sMin \rightarrow \sMax}}$, integrated over the normalised primordial powerlaw $s^{2-3\qP}$). In practise this must be solved numerically, because \eqRef{tauOfGeneralS} for $\sKm$ also depends on $\sMax$. We therefore iterate to find the $\sMax$ value that simultaneously yields $\sKm$ and $\mDiscInitial$. This approach is valid provided the largest bodies are not yet colliding (${\sKm < \sMax}$)\footnote{When the largest bodies start colliding, the disc mass starts to drop according to \eqRef{discMassVsTime}. This occurs in the region below \eqRef{tau0FromM0AndSMax} on \figRef{discMassSMaxAllowedRegionsCollisions}. In this region, our method in \subsecRef{collisionsExtrapolationFromDust} no longer holds.}.

The result is shown as \mbox{Line \ref{eq: massIntegral}} on \figRef{discMassSMaxAllowedRegionsCollisions}. If the current disc obeys the assumed size distribution (\eqRef{sizeDist}), then the initial-disc mass and largest-body size should lie somewhere along this line. The intercept of this line with the dynamical-broadening line from \eqRef{maxMDiscBeforeScatter} yields upper limits on the initial-disc mass and largest-body size. The initial-disc mass would have to have been below ${400\pm100\mEarth}$, and the largest-body radius would have to have been below ${300^{+80}_{-70}\km}$ (${0.3\pm0.1\rPluto}$). Such large bodies would not yet have started colliding, so would still exist today with their primordial size distribution. Since these largest bodies dominate the disc mass, this means that, in this case, the disc would still have its original mass today. Similarly, from the intercept of \eqRef{massIntegral} with \eqRef{tau0FromM0AndSMax}, if the largest primordial bodies still survive today then they must have radii above ${5^{+20}_{-4}\km}$.

\section{Caveats and additional physics}
\label{sec: caveats}

We used $n$-body dynamics and collisional theory to argue that Fomalhaut's debris disc cannot be dominated by primordial Plutos. We now show that this conclusion is robust, by checking various \textit{caveats} and additional physics. \subsecRef{caveatsCollisionsDamping} considers the effects of realistic collisional damping, and \subsecRef{caveatsSizeDist} examines our assumed size distribution. \subsecRef{caveatsOneBigBody} considers whether the largest bodies dominate disc mass, and \subsecRef{caveatsPlutoGrowth} discusses our omission of planetesimal growth. \subsecRefMulti{caveatsGas} and \ref{subsec: caveatsSheperdingPlanets} test whether gas drag or shepherding planets would affect our arguments.

\subsection{Collisional damping by smaller bodies}
\label{subsec: caveatsCollisionsDamping}

Our $n$-body simulations comprise bodies of one size with predominantly inelastic collisions, for which collisional damping is negligible. However, real debris follows a size distribution, so larger bodies could be significantly damped by collisions with smaller bodies. Here we use \cite{Jankovic2024} to show that, whilst realistic collisional damping can sometimes prevent disc broadening, this does not occur in our parameter space of interest. Hence our conclusions would remain valid, even considering more-realistic collisional damping.

\cite{Jankovic2024} show that collisional damping is set by both the collisional timescale and the projectile-to-target-mass ratio $\yC$. Given enough time, damping would eventually become important for a body of a given size if 

\begin{equation}
    \yC = \frac{2 \qDStar}{\vCol^2} \gtrsim 1.
    \label{eq: yCForDamping}
\end{equation}

\noindent Substituting ${\vCol \sim 240 \pm 10 \mPerS}$ from \eqRef{vCol} into \eqRef{yCForDamping}, and using $\qDStar$ for basalt from \eqRef{qDStar}, we find that millimetre dust is always safe from collisional damping in the Fomalhaut disc (${\yC \sim 0.059\pm0.004}$). Conversely, larger objects are not; \eqRef{yCForDamping} shows that, given enough time, collisional damping would eventually become important for bodies larger than ${\sim48 \pm 2\km}$.

However, the second factor is the damping timescale. Large bodies may eventually be susceptible to damping, but this may take so long that it is unimportant in our regime. \cite{Jankovic2024} quantify the damping timescale for a realistic debris-size distribution. Combining their Equations 2 and 5 shows that, for collisional damping to affect the largest bodies within the star lifetime,

\begin{multline}
    \mDisc \gtrsim
                    0.01
                    (4-\alpha)^{-1}
                    \left(\frac{\sMax}{\km}\right)^{4-\alpha}
                    \left[\frac{\rho(\sMax)}{\gPerCmCubed}\right]
                    \left(\frac{\aDisc}{\au}\right)^{5/2}
                    \left(\frac{\daDisc}{\au}\right)\\
                    \times
                    \left(\frac{\mStar}{\mSun}\right)^{-1/2}
                    \left(\frac{\tStar}{\myr}\right)^{-1}
                    \left[\frac{I(\sMax, \alpha)}{\km^{3-\alpha}}\right]^{-1},
    \label{eq: minMDiscToDamp}
\end{multline}

\noindent where $\alpha$ is the size-distribution slope in the region of interest (${n(s) {\rm d}s \propto s^{-\alpha} {\rm d}s}$), and $I(\sMax, \alpha)$ is the integral

\begin{equation}
    I(\sT, \alpha) = \int_{\sMin}^{\sMax} \sP^{-\alpha}(\sP+\sT)^2 \frac{\mP(\sP)\mT(\sT)}{[\mP(\sP)+\mT(\sT)]^2}{\rm d}\sP.
    \label{eq: dampingIntegral}
\end{equation}

\noindent Here subscripts p and t denote properties of the projectile and target bodies respectively, $s$ denotes body radius, $m$ denotes body mass, and ${\sMin}$ is the smallest projectile to consider. We substitute values for Fomalhaut, assuming a size-distribution index for primordial bodies of ${\alpha = 3\qP-2 = 3.61}$ (\eqRef{sizeDist}), calculating body masses using the size-dependent bulk density of \eqRef{bulkDensityVsSize}, and setting ${\sMin = 1\mm}$. The result is the shaded region in the bottom right of \figRef{discMassSMaxAllowedRegions}, which shows only bodies large enough that damping could eventually be important (i.e. ${\yC>1}$).

This analysis demonstrates that, whilst large bodies would eventually be susceptible to collisional damping, the damping timescales are generally much longer than Fomalhaut's age. The only region of our parameter space that is susceptible to damping is debris discs with masses above ${1000\mEarth}$, which is unphysical based on the maximum solid content inheritable from protoplanetary discs \citep{Krivov2021}. We therefore conclude that collisional damping, if it occurs at all, is generally too slow to affect our results. Hence discs that broaden in our $n$-body simulations, often over very short timescales, would not be damped via collisions.

This conclusion is the opposite of \cite{Nesvold2013}, who make a pioneering effort to simultaneously model collisions and $n$-body dynamics. They use the collisional $n$-body code {\sc smack}, and infer that collisions can drive broad discs into narrow, eccentric rings like Fomalhaut. They use a superparticle approach, where each superparticle represents a population of debris bodies with a size distribution. However, they only simulate debris sizes up to ${10\cm}$ and, since {\sc smack} does not include mass segregation, they argue that their method substantially overestimates collisional damping of larger bodies. They also omit debris mass, and hence the effect of scattering. These reasons could explain why we do not reconcile their results with \cite{Jankovic2024}, who show that collisional damping of planetesimals would be negligible in Fomalhaut's disc. We therefore conclude that collisional damping is unlikely to resist scattering-driven broadening, and hence that our conclusions hold, although further work is required to identify exactly why our conclusions differ from \cite{Nesvold2013}.

\subsection{Assumed size distribution}
\label{subsec: caveatsSizeDist}

Our conclusion, that Fomalhaut's disc is not dominated by primordial Plutos, comes from combining dynamical constraints (\eqRef{maxMDiscBeforeScatter}) with disc-mass constraints (\eqRef{minInitialDiscMassFromMassLossRate}). The latter depends on the size distribution around millimetre sizes, because this sets the dust's collisional lifetime and hence the mass-loss rate ${\dot{m}_{\rm disc}}$. However, millimetre dust is firmly in the collisional cascade, so its size distribution is well constrained to lie around ${n(s) \propto s^{-3.5}}$ (e.g. \citealt{Dohnanyi1969, Tanaka1996, Durda1997, OBrien2003, Kobayashi2010, Wyatt2011, Pan2012, Vizgan2022}). We assumed ${s^{-3.67}}$ based on ${\qS=1.89}$ from \cite{Lohne2008}, but changing this to extreme values of ${s^{-3}}$ or ${s^{-4}}$ would only alter our lower bound on initial-disc mass by a factor of 1.7 in either direction (\mbox{Equation 19} in \citealt{Matra2017}). This is not sufficient to let primordial Plutos reside in the Fomalhaut disc (\figRef{discMassSMaxAllowedRegionsCollisions}).


A less-certain parameter is the primordial size distribution, set by ${\qP}$. This is often approximated from planetesimal-formation models (e.g. \citealt{Johansen2015, Simon2016, Simon2017, Abod2019}). However, our main conclusion is robust to ${\qP}$, because it does not directly enter \eqsRef{maxMDiscBeforeScatter} or \ref{eq: minInitialDiscMassFromMassLossRate}. It indirectly comes into \eqRef{maxMDiscBeforeScatter}, by implicitly assuming that the total disc mass is dominated by the largest bodies; however, in \subsecRef{caveatsOneBigBody} we show that \eqsRef{maxMDiscBeforeScatter} is also robust to the size distribution.

The places where ${\qP}$ has the biggest influence is the collisional lifetime of the largest bodies (\eqRef{tau0FromM0AndSMax}), and the extrapolation of dust mass up to total disc mass (\eqRef{massIntegral}). We assume ${\qP=1.87}$ based on \cite{Lohne2008}, which results in ${n(s) \propto s^{-3.61}}$ for large bodies. However, this is uncertain; for example, \cite{Krivov2021} use $s^{-2.8}$ \mbox{(i.e. ${\qP=1.6}$}). To reflect this, we use uncertainties of ${\qP=1.9^{+0.1}_{-0.3}}$ to generate the uncertainties on \figRef{discMassSMaxAllowedRegionsCollisions}. To further test the effect of ${\qP}$, we re-run our collisional analyses with an extreme primordial-size distribution of $s^{-2}$ (i.e. ${\qP=1.3}$). In this case, the collisional-lifetime line (\eqRef{tau0FromM0AndSMax}) moves left on \figRef{discMassSMaxAllowedRegionsCollisions}; the initial-disc mass required for a body to have broken up is reduced by a factor of 5. Also, the dust-extrapolation line (\eqRef{massIntegral}) moves right, such that the the initial-disc mass could be as high as ${1000\mEarth}$ whilst still being consistent with observed dust. In this case the largest-body radius would be below ${220\km}$.

The final parameter is $\qG$, which sets the size distribution in the gravity regime. Whilst this also affects the collisional lifetime of the largest bodies (\eqRef{tau0FromM0AndSMax}), and the extrapolation of dust mass (\eqRef{massIntegral}), it does not affect \eqRef{maxMDiscBeforeScatter} or \eqRef{minInitialDiscMassFromMassLossRate}. This means it has no effect on our conclusion. In summary, since none of the above changes allow Fomalhaut's disc to contain primordial Plutos, our conclusions seem robust to reasonable assumptions about the size distribution.

\subsection{Assumption that the largest bodies dominate disc mass}
\label{subsec: caveatsOneBigBody}

By dividing the disc mass between the largest bodies in our $n$-body simulations, we implicitly assumed that the largest bodies dominate disc mass. This is the standard assumption in debris science, and appears reasonable; for example, Pluto and Eris alone hold ${25\percent}$ of the Kuiper Belt's mass \citep{Pitjeva2018}. It is valid if a disc's size distribution ${n(s) \propto s^{-\alpha}}$ has ${\alpha < 4}$, which is satisfied for debris in a steady state collisional cascade, for which ${\alpha \approx 3.5}$ \citep{Dohnanyi1969}. However, in reality the largest bodies may not yet have entered the cascade, and would still have their primordial size distribution (\subsecRef{collisionsLifetimeOfLargestBodies}). This primordial distribution is uncertain, so it is not clear that the largest bodies in a disc really dominate its mass.

Bodies up to several hundred kilometres in size are thought to form with ${\alpha < 4}$, based on planetesimal-formation models (\citealt{Krivov2021}, and refs. therein). For example, \cite{Krivov2021} use ${2.8 \pm 0.1}$. So if our largest bodies are up to several hundred kilometres then they would still dominate disc mass, in which case our approach would be justified.

However, the upper end of the primordial size distribution may steepen, possibly in the dwarf-planet regime that we are interested in \citep{Schafer2017}. It is therefore possible that the largest body \textit{does not} dominate total disc mass. In this case, it could be that primordial Plutos \textit{are} present in the disc, if the disc mass is dominated by smaller bodies.

We examine this using analytics. \cite{Ida1993} studied the eccentricity growth in a disc comprising numerous small bodies and one embedded large body, where all bodies have mass. Their \mbox{Equation 4.1} includes terms from both small and large bodies, where we neglected large-body terms in \secRef{theoryPrediction}. We now revisit our earlier analysis, this time keeping both their small- and large-body terms, and neglecting the dynamical-friction term of \cite{Ida1993} using the arguments in their \mbox{Section 4}. Using this, we derive the width ratio of two equivalent-mass discs; one comprising only equal-mass bodies, and one where some mass is in one larger body. If $w_\text{M+m}$ is the disc width with the larger body and $w_\text{m}$ that without, then


\begin{equation}
    \frac{w_\text{M+m}}{w_\text{m}} = \left[1 + \frac{M}{\mDisc} \left( \frac{2M}{m} - 1 \right) \right]^{1/4},
    \label{eq: widthRatioOneLargerBody}
\end{equation}

\noindent where ${M}$ and $m$ are the masses of the larger and smaller bodies respectively. Using this, we calculate the effect of introducing a single embedded Pluto into a disc, as shown on \figRef{widthRatiosOneLargeBody}.

\begin{figure}
	\includegraphics[width=8cm]{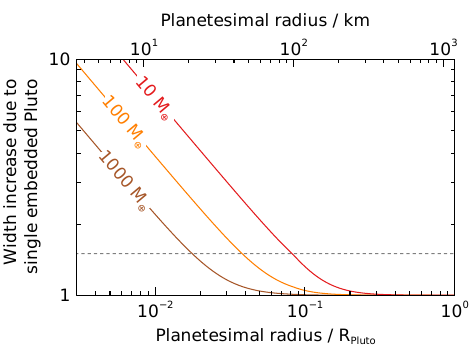}
    \caption{Effect of placing a small fraction of a disc's mass in one Pluto-sized object, where the disc's mass is dominated by smaller objects (\eqRef{widthRatioOneLargerBody}). The horizontal axis shows the radii of the non-Pluto objects, and the vertical axis the ratio of the disc width with and without the Pluto. Three disc masses are shown, as labelled. The plot shows that a Pluto-sized object could be hidden in a disc dominated by smaller bodies without significantly broadening the disc, provided that the smaller bodies have radii larger than ${100\km}$ (for a ${10\mEarth}$ disc) or ${20\km}$ (for a ${1000\mEarth}$ disc). The horizontal dashed line shows a width ratio of 1.5.}
    \label{fig: widthRatiosOneLargeBody}
\end{figure}

The plot shows that one Pluto could be hidden in a disc whose mass is dominated by smaller objects, provided the smaller objects are not too small. Specifically, if Fomalhaut's disc mass is dominated by bodies with radii larger than ${100\km}$ (for a ${10\mEarth}$ disc) or ${20\km}$ (for a ${1000\mEarth}$ disc), then an embedded Pluto would not increase the disc width by more than a factor of 1.5. Therefore, whilst we demonstrate that Fomalhaut's disc cannot be dominated by primordial Plutos, a single Pluto could still be embedded in the disc without significantly broadening it.

This single-Pluto result yields insight into the multi-Pluto case. Additional Plutos would further excite the disc, so \eqRef{widthRatioOneLargerBody} (for a single Pluto) is a lower bound for the multi-Pluto case. If embedded Plutos do not excite each other, then the dynamical effect due to multiple Plutos would be roughly linear with their number. This is quantified by multiplying the factor ${M/\mDisc}$ in \eqRef{widthRatioOneLargerBody} by the number of embedded Plutos, which moves the lines on \figRef{widthRatiosOneLargeBody} upwards. Hence, if the disc contains Plutos but the disc mass is dominated by smaller bodies, then the degree of disc broadening increases with the number of Plutos. Furthermore, in reality multiple embedded Plutos would excite each other, so broaden the disc even more. We therefore conclude that primordial Plutos could be embedded in Fomalhaut's disc, but the disc mass must be dominated by smaller bodies.

\subsection{Omitting planetesimal growth}
\label{subsec: caveatsPlutoGrowth}

We demonstrated that Fomalhaut's primordial disc was not dominated by Plutos, otherwise it would have broadened by now. However, it is possible that planetesimals continued to merge and grow after the protoplanetary disc dissipated, with Plutos forming much more recently. In this case, the younger Plutos might not yet have had time to disrupt the disc, so whilst no Plutos were in the primordial disc, they could be present today.

To test this, we rearrange \eqRef{maxMDiscBeforeScatter} for time $t$, and set ${s=1\rPluto}$ and ${\mDiscInitial = 9 \pm 2 \mEarth}$ (the latter from \eqRef{minInitialDiscMassFromMassLossRate}). This yields the maximum age that Plutos can have without having disrupted the disc, which is ${100 \pm 100\myr}$.
So given the ${440\pm40\myr}$ of Fomalhaut, if Plutos \removed{are present in} \added{dominate} the disc today, then they must have taken at least ${\sim300\myr}$ to form.

Plutos could theoretically form over this timescale; whilst smaller planetesimals are expected to rapidly form via pebble concentration or streaming instability, the subsequent building of Plutos via core accretion can take much longer \citep{Krivov2018SelfStirring}. By combining \mbox{Equations 27} and 41 of \citealt{Kenyon2008}, we find that Plutos could form in Fomalhaut's disc over at least ${300\myr}$, provided that the sub-Plutos were sufficiently unexcited to prevent collisional fragmentation, and the initial planetesimal disc had a surface density less than 5 times the MMSN. This seems reasonable around the ${1.92 \mSun}$ Fomalhaut, so it is possible that, whilst no Plutos existed in the primordial debris disc, they have since formed and will disrupt the disc in future.

In a broader context: collisional models of debris discs typically assume the size distribution evolves with time at intermediate sizes (${\sim\sKm(t)}$), as progressively larger bodies enter the cascade \mbox{(e.g. \citealt{Lohne2008, Krivov2021})}. However, if dwarf-planet formation is also ongoing, then the upper end of the distribution should simultaneously evolve away from a primordial slope as larger bodies grow. There should therefore be a size where the cascade transitions from either destruction- or primordial-dominated to growth-dominated. This could be factored into future models, as it has implications for the total disc mass and the self-stirring level.

\subsection{Gas drag}
\label{subsec: caveatsGas}

Another possibility is that primordial Plutos \textit{are} present in the narrow ring, but that this ring is prevented from spreading by gas drag. This seems unlikely, because the observed-gas content of the Fomalhaut ring is low, especially compared to other A stars \citep{Matra2017, Hughes2018}. To test this, we consider two possibilities; either gas confines planetesimals to a narrow ring, or gas confines observed millimetre dust, whilst unseen planetesimals have a wider distribution.

To test the first possibility, we consider planetesimals in a narrow, eccentric gas disc. These planetesimals would interact with gas in the Stokes drag regime, with acceleration ${\ddot{\vect{r}}} \propto -|\vect{\vRel}| \vect{\vRel}$ (where $\vect{\vRel}$ is the planetesimal's velocity relative to the gas). \cite{Beauge2010} show that, in a non-precessing eccentric disc, gas eventually drives such planetesimals onto narrow, eccentric orbits matching the gas. This occurs over the proper-eccentricity damping timescale. Combining their \mbox{Equations 4}, 18 and 19, we find that a planetesimal's proper eccentricity $\eP$ is damped by gas on a timescale

\begin{equation}
    \tau_{\eP} \equiv \frac{\eP}{|\dot{e}_{\rm p}|} = \frac{32 s \rho(s)}{3\pi \eP C_{\rm D} \rhoGas} \sqrt{\frac{a}{G \mStar}},
    \label{eq: gasDampTimescalePlanetesimal}
\end{equation}

\noindent where $\rhoGas$ is the gas volume density and $C_{\rm D}$ is a drag coefficient (${C_{\rm D} = 0.44}$ for $1\km$ planetesimals; \citealt{Adachi1976, Weidenschilling1977}).

Applying this to Fomalhaut: the observed CO gas mass has an upper limit ${\mGas < 4.9 \times 10^{-4}\mEarth}$ \citep{Matra2015}. If the gas is co-located with the dust, then its density is roughly ${\rhoGas \approx \mGas /[2 \pi^2 \aDisc (\daDisc/2)^2] < 8.7\times 10^{-21}\gPerCmCubed}$. If no other gas is present, then \eqRef{gasDampTimescalePlanetesimal} shows that a ${1\km}$ planetesimal would take at least ${1.1\times 10^{14}\yr}$ to damp into a narrow ring, which is far longer than Fomalhaut's ${4.4\times10^{8}\yr}$ age. These timescales are even longer for larger planetesimals. There could be more gas in other, undetected species, but \eqRef{gasDampTimescalePlanetesimal} shows that at least ${100\mEarth}$ of gas would be needed to damp planetesimals within Fomalhaut's age. Assuming the unseen gas is predominantly H$_2$, this requires a \mbox{CO/H$_2$} abundance ratio of at most ${10^{-7}}$, which is much lower than typical ISM values of ${10^{-4}}$ (e.g. \citealt{Kospal2013, Matra2015}). Hence it seems unlikely that Fomalhaut's planetesimals are confined to a narrow ring by gas.

The second possibility is that the planetesimal belt is wider than the observed millimetre belt, but that the millimetre dust is confined by a narrow ring of gas. For plausible gas quantities in the Fomalhaut disc, the mean-free path of gas molecules is much larger than ${1\mm}$, regardless of gas species. This means that dust interacts with gas in the Epstein drag regime, with an acceleration of either ${\ddot{\vect{r}}} \propto -\vect{\vRel}$ or ${-|\vect{\vRel}| \vect{\vRel}}$ if the grain is subsonic or supersonic respectively (e.g. \citealt{Kwok1975, Armitage2010, Pearce2020}). For simplicity we assume the grain is supersonic, to drive maximal damping. In this case ${\ddot{\vect{r}} \propto -|\vect{\vRel}| \vect{\vRel}}$ and we can again use the \cite{Beauge2010} formalism, only the prefactor ${\mathcal{C}}$ in their \mbox{Equation 4} now becomes ${\rhoGas/[s \rho(s)]}$. This means that \eqRef{gasDampTimescalePlanetesimal} is reduced by a factor ${8/(3C_{\rm D})}$ when applied to supersonic dust.

Substituting numbers for Fomalhaut: if only CO gas is present, then millimetre grains would be damped into a narrow ring over ${17\myr}$. However, these grains would be destroyed in collisions before damping could complete, over a ${1\myr}$ timescale \citep{Pearce2021}. Yet this raises the possibility that realistic gas masses \textit{could} confine millimetre grains; if observed CO constitutes at most ${10\percent}$ of the total gas mass, then dust could be confined by gas. However, we do not consider this realistic, for two reasons. First, the ring is relatively sharp in ALMA images \citep{MacGregor2017}, with very little millimetre emission either side. If planetesimals actually had a broader distribution, then we would expect some millimetre emission either side of the ring from dust released in planetesimal collisions. This is not observed. Second, gas would also perturb ${\sim10\um}$ grains as they migrated inwards from the belt through Poynting–Robertson drag. Such grains are seen by the \textit{James Webb Space Telescope} (\textit{JWST}), yet their profile is well-explained by a gas-free model \citep{Gaspar2023, Sommer2025}. We therefore find it unlikely that gas maintains Fomalhaut's narrow ring in the face of dynamical spreading, by either confining planetesimals or dust, so we do not believe that gas affects our overall conclusions.

\subsection{Shepherding planets}
\label{subsec: caveatsSheperdingPlanets}

A final possibility is that the Fomalhaut ring would broaden by itself, but is prevented from doing so by shepherding planets. In this mechanism, planets exert secular torques that repel the ring, so the combined effect of two planets either side of the ring is to confine planetesimals to a narrow range of semimajor axes (e.g. \citealt{Murray1999}).

\cite{Goldreich1979} first proposed shepherding to explain narrow rings of Uranus, which would otherwise broaden through collisions. They argue that a circular, massless, collisionless ring, shepherded by a pair of equal-mass planets at equal distances either side of the ring, would narrow over a timescale

\begin{equation}
    \tShep \sim \frac{\Omega_{\rm disc}^3 \aDisc x^5}{f_1 G^2 \mPlt},
    \label{eq: shepherdingTimescale}
\end{equation}

\noindent where $\Omega_{\rm disc}$ is the disc angular velocity, $x$ the disc-planet separation, $\mPlt$ the planet mass, and $f_1$ a constant of order 1. They argue that, for a ring in steady state, the shepherding timescale $\tShep$ should equal the ring's broadening timescale, which in Uranus' case is the diffusion timescale due to collisions.

For Fomalhaut, the broadening mechanism is self scattering, and the broadening timescale can be estimated from our $n$-body simulations of non-shepherded rings (\secRef{nBody}). For shepherding planets to counteract scattering and keep the ring narrow, the shepherding timescale from \eqRef{shepherdingTimescale} should be at most the broadening timescale from our earlier simulations.

To test this, we run new $n$-body simulations of the Fomalhaut disc plus two shepherding planets. The planets are equal mass, one on either side of the disc, and on coplanar orbits apsidally aligned with the disc. We consider the disc parameters from \figRef{simPlotHighMass}: a ${30\mEarth}$ disc comprising bodies of radius ${1.9\rPluto}$, which our $n$-body simulations show would broaden over ${\sim20\myr}$ if not shepherded. We run this simulation three more times, each time including two equal-mass planets of mass 0.7, 0.2 or ${0.07\mJup}$ (200, 70 and ${20\mEarth}$ respectively), where ${0.7 \mJup}$ is the $3\sigma$ upper limit near the disc from \textit{JWST} NIRCam \citep{Ygouf2024}. In each case the planets' semimajor axes are calculated using \eqRef{shepherdingTimescale}, with $\tShep$ set to ${20\myr}$. We set the inner planet's eccentricity to 0.2. We calculate the outer planet's eccentricity such that the forced eccentricity of a test particle at semimajor axis $\aDisc$ equals the observed disc's forced eccentricity $\eForced$, using the two-planet secular analysis in \cite{Murray1999}. Using these constraints, for planet masses of 0.7, 0.2 and ${0.07\mJup}$ we use inner-planet semimajor axes of 78.9, 102 and ${113\au}$ respectively,
and the corresponding outer-planet semimajor axes are 194, 171 and ${159\au}$.
The outer planet's eccentricities are 0.12, 0.08 and 0.06
respectively. All planet pairs are separated by more than ${2\sqrt{3}}$ mutual Hill radii, as required for dynamical stability \citep{Chambers1996}, and are initially more than 3 Hill radii from the disc edge to prevent disc scattering (e.g. \citealt{Gladman1993, Ida2000, Kirsh2009, Pearce2024Edges}).

We find that the shepherded discs still broaden, and even more so than the non-shepherded case. There are two possible reasons for this. First, the planets oscillate in eccentricity due to mutual secular interactions (e.g. \citealt{Murray1999}). As the planets evolve, the forced eccentricity at the disc location changes, so the disc shape also changes. Second, the planets eventually start scattering disc material, which drives the planets towards the disc and thus destabilises the system. This scattering starts when the planets first encounter debris, either as it leaves the disc through debris self-scattering, or as debris orbits change via the above secular interactions. Either way, when the inner planet encounters external debris it starts migrating outwards, and \textit{vice versa} for the outer planet. Hence the planets converge and the intervening disc is broadened. We repeated the shepherding simulations with several other parameters; we varied the disc mass by a factor of 3, the debris-body size by a factor of 2, and the planet locations by ${30\percent}$ in either direction. In all cases, the discs ended up broader than without shepherding.

We conclude that shepherding of Fomalhaut's eccentric disc, if it can occur, would require a delicate balance of system parameters. We were unable to identify a setup where shepherding truly increased the dynamical stability of the system, and the issue is increasingly acute for more massive discs. We note that \cite{Boley2012} did not encounter this problem in their shepherding simulations, because they simulated planets interacting with a massless disc. Whilst we do not conduct a thorough exploration, our simulations suggest that shepherding is difficult to achieve in practise. We therefore believe it unlikely that Fomalhaut's disc contains primordial Plutos, which is prevented from broadening by unseen shepherding planets.

\section{Discussion: wider impact}
\label{sec: discussionContext}

We used a new method to show that Fomalhaut's disc mass cannot be dominated by primordial Plutos and that, if Plutos are present, they must have formed recently. This study has wider implications for several areas of debris science, beyond Fomalhaut alone.

First, our conclusion backs up the argument of \cite{Krivov2021}, that `planetesimals are born small' in the brightest debris discs. They argue that the largest bodies must be small, because extrapolating observed dust up to unseen Plutos would mean that the brightest debris discs have masses above ${1000\mEarth}$, which would violate our understanding of protoplanetary discs \citep{Krivov2018PltmlFormation}. Their argument is solely based on collision theory, but here we reach a similar conclusion using dynamics. Whilst \cite{Krivov2021} argue that the dominating bodies may be just ${\sim10\km}$ or even ${\sim1\km}$ in size, which is unfortunately well below our dynamical limits, we nonetheless agree that the Fomalhaut disc cannot be dominated by primordial Plutos. This means there are now two independent arguments that Plutos are absent from at least some bright debris discs, unless the size distribution turns over sharply at the upper end, or Plutos formed recently.

Second, a major problem in debris science is that we do not know how massive debris discs are. This is because we cannot observe large bodies, which are expected to dominate disc mass. Our paper offers a new technique to dynamically constrain the masses of narrow debris discs. It can easily be applied to other systems, such as ${\HD53143}$, ${\HD181327}$, and ${\HD202628}$, and we intend to incorporate this into a wider study in the future. Our analytic methods, such as \eqRef{maxMDiscBeforeScatter}, can be applied without running $n$-body simulations, to constrain disc mass and largest-body size. Note that a disc need not be globally eccentric for our methods to apply; the only requirement is that it is narrow. This manifests as the ${\daDisc^5}$ term in \eqRef{maxMDiscBeforeScatter}; if the disc is too wide, then our constraints become significantly weaker.

Our technique adds to a small but growing arsenal of methods to weigh debris discs, which include: using vertical thickness or edge steepness to constrain the masses of stirring bodies (e.g. \citealt{Ida1993, Matra2019, Marino2021, ImazBlanco2023}), calculating the masses of eccentric discs required to resist planetary shearing \citep{Pearce2023}, using nearby planets to constrain the masses of gapped discs from either secular effects \citep{Pearce2015HD107146, Sefilian2021} or planet migration \citep{Friebe2022, Booth2023}, and using inclined planets to infer disc masses from vertical structure \citep{Poblete2023, Sefilian2025}. No single technique can robustly weigh all types of debris disc, so for now, insights must come from targeted constraints using specific methods for specific systems. However, our list of techniques is growing, and expanding our knowledge of debris-disc masses is the thrust of several ongoing works.

\section{Conclusions}
\label{sec: conclusions}

We use dynamical models and $n$-body simulations to constrain the largest bodies in the Fomalhaut debris disc. If the bodies were too large, or the disc too massive, then bodies would scatter each other, resulting in a significantly broader disc than observed today. Our main conclusions are:

\begin{enumerate}
    \item Primordial Plutos do not dominate the mass of Fomalhaut's debris disc.
    \item The only way that Plutos could dominate is if they formed recently, within the last ${150\myr}$ (compared to the ${440\myr}$ age of Fomalhaut).
    \item The only way for primordial Plutos to exist in the disc today is if ${<100\km}$ bodies dominate the disc mass.
\end{enumerate}

\noindent We investigate many alternative scenarios to maintain the narrow disc, including shepherding planets, collisional damping and gas drag, but find that the above conclusions are robust.

Our paper presents a new method for dynamically constraining debris-disc masses and largest-body sizes, both of which are key unknowns. It provides a new, independent argument that the brightest discs cannot be dominated by large primordial bodies. This provides independent support for \cite{Krivov2021} who, using collisional models, argue that the largest bodies in the brightest debris discs are much smaller than Pluto.

\section*{Acknowledgements}

We thank \added{Sebasti\'{a}n Marino}\removed{the anonymous referee} for \added{his}\removed{their} thorough and insightful review, which significantly improved the science and clarity of the paper. \added{We also thank Antranik Sefilian for a clarification in \appRef{secularSimulations}}. TDP is supported by a UKRI Stephen Hawking Fellowship and a Warwick Prize Fellowship, the latter made possible by a generous philanthropic donation. TDP and AVK also acknowledge past support from DFG KR 2164/13-2, which led to the discovery of the `debris-disc mass problem'. 

\section*{Data Availability}

The data underlying this article will be shared upon reasonable request to the corresponding author.


\bibliographystyle{mnras}
\bibliography{bib} 

\appendix

\section{Bulk densities of Solar System bodies}
\label{app: densities}

We make an approximate relation between debris-body size and bulk density, based on solid bodies in the Solar System. We take sizes and bulk densities from literature\footnote{References for Solar System bodies: JPL Solar System Dynamics (\url{https://ssd.jpl.nasa.gov/planets/phys_par.html}); \citet{Anderson1987, Konopliv1999, Jacobson2006, Brown2007, Jacobson2009, Ragozzine2009, Thomas2010, Brown2013, Jacobson2014, Lockwood2014, Brozovic2015, Park2016, Patzold2016, Archinal2018, Ernst2023}.}, and plot these on \figRef{ssBulkDensities}. We then fit (by eye) a simple trend relating bulk density to body size. This is \eqRef{bulkDensityVsSize}, shown by the solid line on \figRef{ssBulkDensities}. Whilst different Solar System bodies have different compositions, so are not expected to follow a single trend, this approximation is good enough for our purposes. 

\begin{figure}
	\includegraphics[width=8cm]{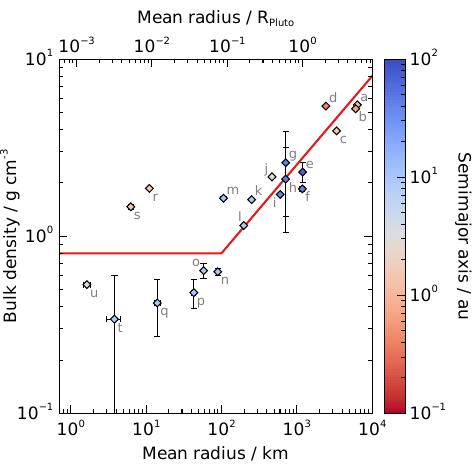}
    \caption{Red line: simple approximation for the bulk densities of solid bodies versus their size, as adopted in our $n$-body simulations (\eqRef{bulkDensityVsSize}). Points show various Solar System objects for comparison:
    $^\text{a}$Earth,
    $^\text{b}$Venus, 
    $^\text{c}$Mars, 
    $^\text{d}$Mercury, 
    $^\text{e}$Eris, 
    $^\text{f}$Pluto, 
    $^\text{g}$Haumea, 
    $^\text{h}$Makemake, 
    $^\text{i}$Charon, 
    $^\text{j}$Ceres, 
    $^\text{k}$Enceladus, 
    $^\text{l}$Mimas, 
    $^\text{m}$Phoebe, 
    $^\text{n}$Janus, 
    $^\text{o}$Epimetheus, 
    $^\text{p}$Prometheus, 
    $^\text{q}$Pan, 
    $^\text{r}$Phobos, 
    $^\text{s}$Deimos, 
    $^\text{t}$Daphnis, 
    $^\text{u}$Comet 67P.
    For any data without literature uncertainties, we assume uncertainties of ${50\percent}$. Colours denote heliocentric semimajor axis; for satellites, this is that of the parent body.}
    \label{fig: ssBulkDensities}
\end{figure}

\section{Secular-ring simulations}
\label{app: secularSimulations}

The $n$-body simulations in \secRef{nBody} require prohibitively long computation times for large numbers of particles. We therefore make two assumptions to reduce computational requirements:

\begin{enumerate}

    \item That narrow discs comprised of small bodies would remain narrow. This lets us omit $n$-body simulations with very large numbers of particles (lower region of \figRef{discMassSMaxAllowedRegions}).
    
    \item That broadened discs would not damp themselves back into narrow rings through self-gravity. This lets us terminate $n$-body simulations early for discs that quickly broaden.
    
\end{enumerate}

\noindent In this section we verify that both assumptions are valid.

Gravitational interactions can be split into 3 categories: scattering, secular, and mean-motion-resonance interactions. Scattering interactions are close approaches that modify orbits over relatively short timescales; these are what cause discs in our $n$-body simulations to broaden. Conversely, secular interactions occur over much longer timescales.

Full $n$-body simulations include all 3 interactions. As a first test, we continue several $n$-body simulations for long after the discs significantly broaden, and do not detect any later narrowing. These test simulations cover a significant region of the parameter space, and demonstrate that scattering is unlikely to narrow broadened discs.

As a second test we run a different type of simulation, focussing on secular interactions. Secular dominates over scattering if particles are very small (\mbox{Section 4.1.3} in \citealt{Costa2024}), or if the interaction occurs over a long time. We therefore use a secular integrator to test our two assumptions above. We describe the techniques below.

\subsection{Secular integrator}

Secular interactions occur on timescales much longer than orbital periods. A disc undergoing secular interactions can therefore be modelled as being composed of ring-like sub belts, where each ring gravitationally interacts with others, causing them to change shape and orientation. To model this, we implement a numerical integrator based on the secular theory below.

To lowest order in eccentricities and inclinations, the orbit-averaged gravitational influence of a set of $N-1$ perturbers on a particle can be expressed via the disturbing potential

\begin{equation}
    \mathcal{R}_{\rm i} = n_{\rm i} a_{\rm i}^2 \sum_{\rm j\neq i}\left[\frac{1}{2}A_{\rm ii}(h_{\rm i}^2 + k_{\rm i}^2) +  A_{\rm ij} (h_{\rm i} h_{\rm j} + k_{\rm i} k_{\rm j})\right]
\end{equation}

\noindent (e.g. \citealt{Murray1999}). Here the subscript ${\rm i}$ denotes properties of the particle, ${\rm j}$ the perturbers, ${h \equiv e \sin \varpi}$ and ${k \equiv e \cos \varpi}$ where $e$ and $\varpi$ are eccentricity and longitude of ascending note respectively, $a$ is the semimajor axis, ${n_i \equiv \sqrt{G(\mStar + m_i) a_i^{-3}}}$ is the mean motion, and $m$ is the mass. The values $A_{\rm ii}$ and $A_{\rm ij}$ are elements of the interaction matrix $\mat{A}$, where

\begin{equation}
    A_{\rm ij} = 
    \begin{cases}
      \sum_{{\rm k\neq i}}\limits \frac{n_{\rm j} m_{\rm k}}{M_* + m_{\rm j}} \frac{\fHahn(\alpha_{\rm ik})}{4} & \text{if}\ {\rm i=j}, \\
      -\frac{n_{\rm i} m_{\rm j}}{M_* + m_{\rm i}} \frac{\gHahn(\alpha_{\rm ij})}{4} & \text{if}\ {\rm i\neq j}.
    \end{cases}
\end{equation}

\noindent Here the functions ${f = f(\alpha_{\rm ij},H)}$ and ${g = g(\alpha_{\rm ij}, H)}$ represent the softened Laplace coefficients as introduced by \cite{Hahn2003}, where $\alpha_{\rm ij} \equiv a_{\rm i}/a_{\rm j}$ for all combinations of $a_{\rm i}$ and $a_{\rm j}$. The softening parameter $H$ can be associated with the relative vertical thickness of a disc, $H \approx \Delta z/r$, accounting for the fact that the local interactions between neighbouring orbits are fundamentally different from that in an infinitely thin disc (see \added{\citealt{Sefilian2019_Softening}} for a review). We set ${H=0.06}$ for narrow belts and ${H=0.15}$ for wide belts, although these choices have no significant influence on our conclusions.

In vector notation, the resulting secular changes to the orbital elements of all interacting particles follow from Lagrange's planetary equations

\begin{equation}
    \dot{\vect{h}} = \mat{A} \vect{k} \quad \text{and} \quad \dot{\vect{k}} = -\mat{A} \vect{h},
    \label{eq: hk_change_1}
\end{equation}

\noindent where ${\vect{h} \equiv (h_1, ..., h_{\rm N})}$ and ${\vect{k} \equiv (k_1, ..., k_{\rm N})}$ \citep{Murray1999}. Up to this point our implementation of the $N$-ring approach is identical to the models presented by \cite{Hahn2003} and \cite{Sefilian2023}, the results of which can be reproduced exactly. Our approach to a numerical solution differs only in that we first decouple \eqsRef{hk_change_1} into separate sets of undamped harmonic oscillators for $h$ and $k$ by a second application of $\mat{A}$:

\begin{equation}\label{eq:hk_change_2}
    \ddot{\vect{h}} = -\mat{A}^2 \vect{h} \quad \text{and} \quad \ddot{\vect{k}} = -\mat{A}^2 \vect{k}.
\end{equation}

\noindent The eigenvalues and eigenvectors to $-\mat{A}^2$ are then sought with the C++ {\sc eigen} library, and when combined with initial conditions ${h_{\rm i}(t = 0)}$ and ${k_{\rm i}(t = 0)}$, result in time-dependent solutions

\begin{equation}
  h_{\rm i} = \sum_{\rm j=1}^N b_{\rm ij} \sin(c_{\rm j} t + d_{\rm j}) \quad \text{and} \quad
  k_{\rm i} = \sum_{\rm j =1}^N b_{\rm ij} \cos(c_{\rm j} t + d_{\rm j})
  \label{eq: secHAndKEvo}
\end{equation}

\noindent with amplitudes $b_{\rm ij}$, frequencies $c_{\rm j}$ and phase shifts $d_{\rm j}$ \added{\citep[cf.][Chapter 7]{Murray1999}}. This solution is valid for an arbitrary time $t$, and the solution is bounded unless additional perturbers induce secular resonances \added{\citep[e.\,g.][]{Sefilian2021,Sefilian2023}}, where one or several of the amplitudes $b_{\rm ij}$ diverge\removed{\citep{Sefilian2023}}.

We implement the above theory to model the evolution of 1000 massive sub belts, where each sub belt represents the orbits of bodies in a disc. For a given setup, we calculate the system architecture due to secular interactions at time $t$ using \eqRef{secHAndKEvo}. We use this to make the two tests below.


\begin{figure*}
	\includegraphics[width=17cm]{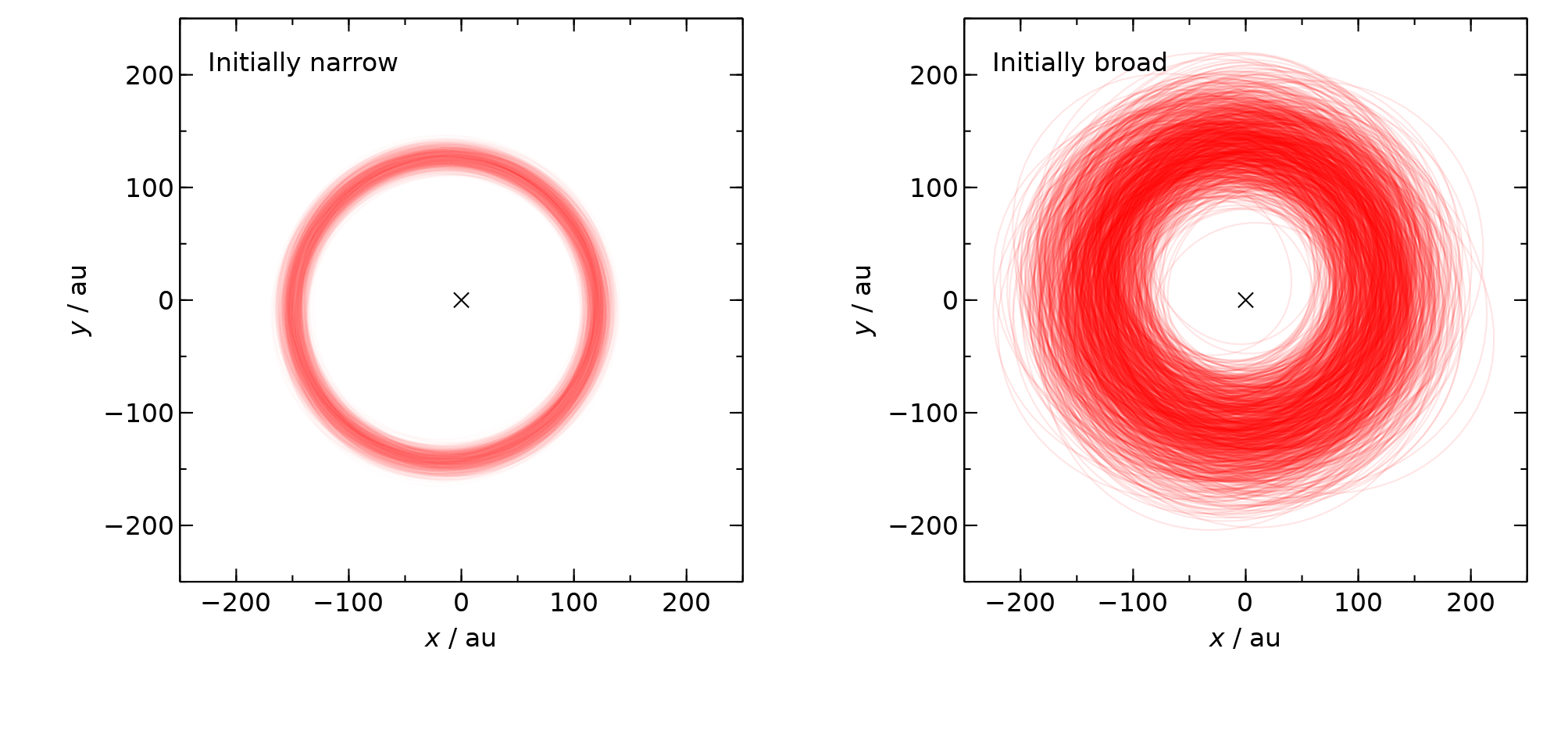}
    \caption{Secular simulations of two ${3000\mEarth}$ debris discs, each shown at ${440\myr}$ (\appRef{secularSimulations}). The cross marks the star, and each red ring is an interacting sub-belt. Left panel: an initially narrow, eccentric disc, with the same initial setup as the Fomalhaut disc (\subsecRef{nBodySetup}). Right panel: an initially broad, eccentric disc. Each disc maintains its initial global eccentricity and width throughout the simulation, and the results are qualitatively identical for all tested disc masses (0.1 to ${3000\mEarth}$). This demonstrates that secular effects would not cause our narrow discs to broaden, nor our broadened discs to narrow.}
    \label{fig: secularSims}
\end{figure*}

\subsection{Conclusion 1: narrow secular discs do not broaden}
\label{appSubsec: secularSimulations_narrow}

Our first test is whether a narrow disc comprising an $\sim$infinite number of $\sim$infinitesimally small bodies would broaden over time. Such a disc would be entirely secular. We model this using the secular code, considering 12 discs with logarithmically spaced masses between 0.1 and ${3000 \mEarth}$. Each disc comprises \mbox{1000 sub-belts}, set up with the same parameters as the initial orbits in our $n$-body simulations (\subsecRef{nBodySetup}). The secular simulations are run for the ${440\myr}$ age of Fomalhaut, and an example simulation with the highest disc mass is shown on the left panel of \figRef{secularSims}.

None of these stimulations show any significant broadening, and the mean free and forced eccentricities remain very close to their initial values. We can therefore conclude that, for our parameter space, secular effects do not cause narrow discs to broaden. This validates our decision to omit $n$-body simulations of discs with large numbers of small particles; these discs would not broaden over time. These discs do exhibit global precession, but they maintain their global eccentricity and width.

\subsection{Conclusion 2: broadened secular discs do not narrow}
\label{appSubsec: secularSimulations_broad}

Our second test is whether a broadened disc would damp itself back into a narrow ring through secular effects. We again use the secular code to model 12 discs with logarithmically spaced masses between 0.1 and ${3000 \mEarth}$. We use a similar setup as in \appSubsecRef{secularSimulations_narrow}, except with a much broader disc (semimajor axes spanning 70 to ${200\au}$), with orbits that are much more excited (proper eccentricities spanning ${0 \leq \eFree \leq 0.3}$). This corresponds to the state of $n$-body simulations where the disc has broadened to ${\sim5}$ times its initial value. The secular simulations are again run for ${440\myr}$, and an example for the highest disc mass is shown on the right  panel of \figRef{secularSims}.

Again, no simulations produce a significant change in the global disc morphology. This confirms that, for our parameter space, a broadened debris disc will not become narrow through secular effects. This validates our decision to terminate $n$-body simulations early if the disc significantly broadens, because such a broad disc would remain broad.

\section{Tree-gravity test}
\label{app: treeCodeTests}

To verify our {\sc rebound} tree-gravity simulations, we compare two batches of test simulations, one using tree gravity and the other simple $n$-body gravity. We run 12 setups of the Fomalhaut disc, each with 300 debris bodies, with disc masses logarithmically spaced between 0.03 and ${10^4\mEarth}$. For each setup we run two simulations, one using tree and one using $n$-body gravity, and compare the resulting disc widths (as measured using the method in \subsecRef{nBodyAnalysis}). The results are shown on \figRef{treeVsNBodyTest}.

\begin{figure}
	\includegraphics[width=8cm]{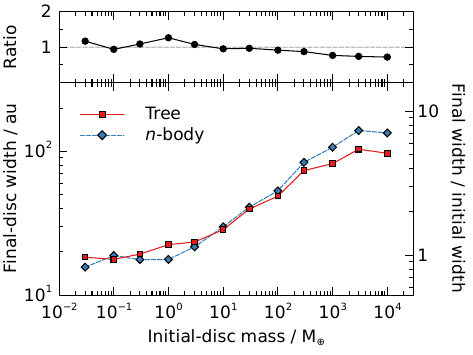}
    \caption{Comparison of simulations using tree gravity to those using simple $n$-body gravity (\appRef{treeCodeTests}). Bottom panel: disc widths measured at the end of each simulation, for tree gravity (squares) and $n$-body gravity (diamonds). Top panel: ratio of disc widths from tree gravity to $n$-body gravity. Each simulation has 300 debris bodies, and the setups are the same for each gravity type. Each pair of disc widths are within ${25\percent}$ of each other, validating our tree-gravity setup. The turnover at high disc masses is due to our method struggling to fit very broadened discs, whose profiles deviate strongly from Gaussian.}
    \label{fig: treeVsNBodyTest}
\end{figure}

The tree and $n$-body simulations show very similar behaviour, with both predicting little disc broadening for disc masses below ${\sim10\mEarth}$ and then increasing for higher masses. At these high masses, tree gravity yields slightly narrower discs than $n$-body gravity, but the results deviate by less than ${25\percent}$, which is not enough to affect our analyses. We therefore conclude that our tree-gravity setup is valid.

\section{Collisional functions}
\label{app: collisionFunctions}

The functions ${f(e,i)}$ and ${G(q, s, r)}$ in \eqRef{tau0FromM0AndSMax} are defined in \cite{Lohne2008} as

\begin{equation}
    f(e, i) = \sqrt{\frac{5}{4} e^2 + i^2}
    \label{eq: fInCollisionModel}
\end{equation}

\noindent and

\begin{multline}
    G(q, s, r) \equiv \left[\xC(s, r)^{5-3q}-\left(\frac{\sMax}{s}\right)^{5-3q} \right]\\
    + 2\frac{q-5/3}{q-4/3}\left[\xC(s, r)^{4-3q}-\left(\frac{\sMax}{s}\right)^{4-3q} \right]\\
    + \frac{q-5/3}{q-1}\left[\xC(s, r)^{3-3q}-\left(\frac{\sMax}{s}\right)^{3-3q} \right],
    \label{eq: GInCollisionModel}
\end{multline}

\noindent where 

\begin{equation}
    \xC(s,r) \equiv \left[\frac{2 \qDStar(s,r) r}{f^2(e, I) \mathcal{G} m_*}\right]^{1/3}.
    \label{eq: XcInCollisionModel}
\end{equation}

\bsp	
\label{lastpage}
\end{document}